\documentclass[sigconf]{acmart}

\usepackage{subcaption}

\usepackage[T1]{fontenc}

\usepackage[utf8]{inputenc}
\usepackage{microtype}
\usepackage{multirow}

\usepackage{xcolor}

\definecolor{promptblue}{RGB}{210,230,250}
\definecolor{promptgreen}{RGB}{210,250,210}
\definecolor{promptyellow}{RGB}{250,250,210}
\definecolor{promptred}{RGB}{250,210,210}
\definecolor{promptpurple}{RGB}{230,210,250}
\definecolor{promptorange}{RGB}{255,228,196}
\definecolor{promptcyan}{RGB}{224,255,255}
\definecolor{promptgray}{RGB}{230,230,230}

\AtBeginDocument{%
  }

\setcopyright{acmlicensed}
\copyrightyear{2026}
\acmYear{2026}
\setcopyright{cc}
\setcctype{by}
\acmConference[CHI '26]{Proceedings of the 2026 CHI Conference on Human Factors in Computing Systems}{April 13--17, 2026}{Barcelona, Spain}
\acmBooktitle{Proceedings of the 2026 CHI Conference on Human Factors in Computing Systems (CHI '26), April 13--17, 2026, Barcelona, Spain}
\acmPrice{}
\acmDOI{10.1145/3772318.3791236}
\acmISBN{979-8-4007-2278-3/2026/04}

\begin{document}

\title[AI Meets Mathematics Education]{AI Meets Mathematics Education: A Case Study on Supporting an Instructor in a Large Mathematics Class with Context-Aware AI}

\author{J\'er\'emy Valentin Barghorn}
\authornote{Both authors contributed equally to this research.}
\affiliation{%
  \institution{\'Ecole polytechnique f\'ed\'erale de Lausanne}
  \country{Switzerland}
}

\author{Anna Sotnikova}
\authornotemark[1]
\affiliation{%
  \institution{\'Ecole polytechnique f\'ed\'erale de Lausanne}
  \country{Switzerland}
}
\email{anna.sotnikova@epfl.ch}

\author{Sacha Friedli}
\affiliation{%
  \institution{\'Ecole polytechnique f\'ed\'erale de Lausanne}
  \country{Switzerland}
}

\author{Antoine Bosselut}
\affiliation{%
  \institution{\'Ecole polytechnique f\'ed\'erale de Lausanne}
  \country{Switzerland}
}

\renewcommand{\shortauthors}{Barghorn, Sotnikova et al.}

\begin{abstract}
Large-enrollment university courses face persistent challenges in providing timely and scalable instructional support. While generative AI holds promise, its effective use depends on reliability and pedagogical alignment. We present a human-centered case study of AI-assisted support in a Calculus I course, implemented in close collaboration with the course instructor. We developed a system to answer students’ questions on a discussion forum, fine-tuning a lightweight language model on 2,588 historical student–instructor interactions. The model achieved 75.3\% accuracy on a benchmark of 150 representative questions annotated by five instructors, and in 36\% of cases, its responses were rated equal to or better than instructor answers. Post-deployment student survey (N = 105) indicated that students valued the alignment of the responses with the course materials and their immediate availability, while still relying on the instructor verification for trust. We highlight the importance of hybrid human–AI workflows for safe and effective course support.
\end{abstract}



\begin{teaserfigure}
      \centering
  \includegraphics[width=\textwidth]{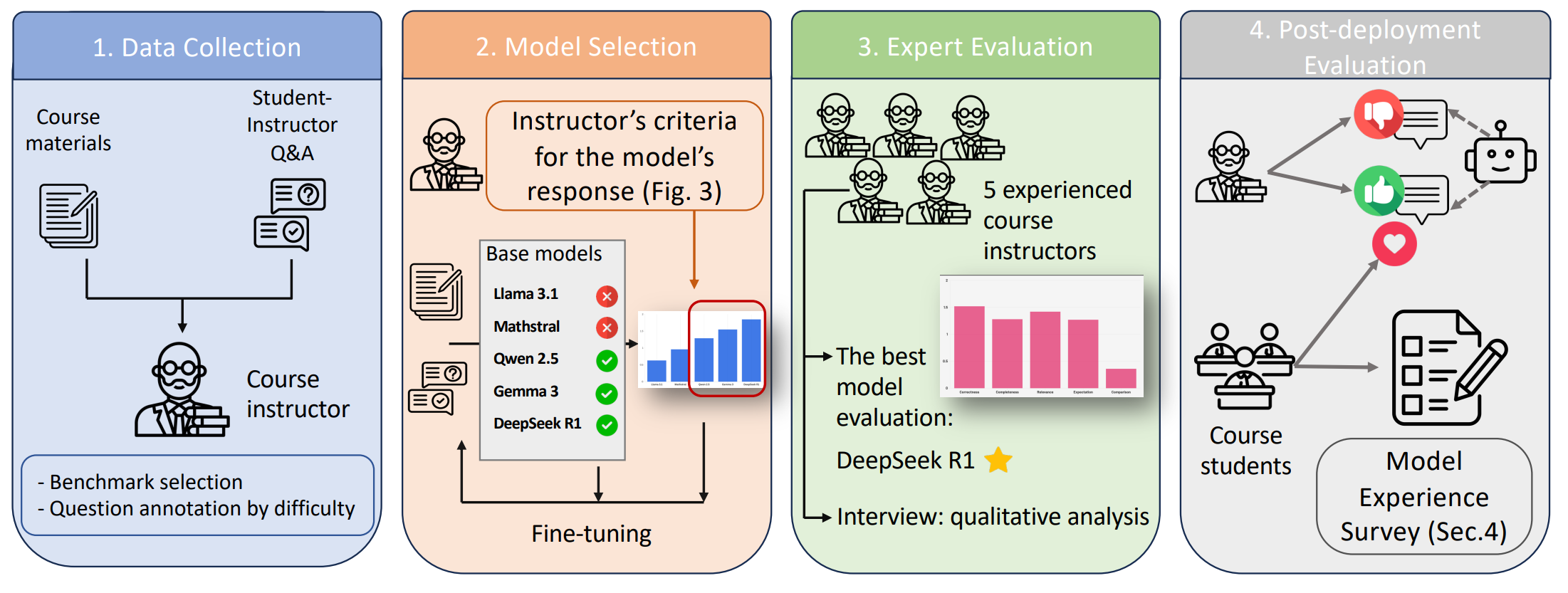}
  \caption{Overview of our human-guided research process: (1) collecting and annotating student–instructor Q\&A pairs, (2) instructor-led model fine-tuning and evaluation, (3) expert evaluation by five Calculus instructors through surveys and interviews, and (4) post-deployment evaluation using student surveys and instructors' review of model responses.}
  \Description{}
  \label{fig:layout_of_work}
\end{teaserfigure}


\maketitle

\section{Introduction}
Artificial intelligence (AI) is increasingly adopted in education to support tasks such as automated grading, tutoring, and content generation~\citep{Miroyan, LeeHaeJin, Bodonhelyi_2025, Yucheng, Ferreira}. With both students and instructors incorporating these tools into their daily practices, traditional learning environments are undergoing rapid transformation~\citep{freeman2025student, du2024llmchem, usdoe2023ai, Cai, shaer2024aiaugmentedbrainwritinginvestigatinguse}. Alongside their potential to enhance education, generative AI systems also introduce risks, including inaccurate or misleading feedback~\citep{Borges2024, Jukiewicz, Qian2025}, bias and discrimination~\citep{Monib2024, chinta2024fairaiednavigatingfairnessbias, Li_2025}, and misalignment with pedagogical goals~\citep{Aad, Liu_Liyuan}. In response, a growing body of research emphasizes human-centered approaches to the design and evaluation of educational AI, incorporating instructors and learners into the development process. These efforts move beyond abstract discussions to address practical use cases, such as AI-assisted course support, where human oversight and iterative feedback are crucial.

In this work, we also adopt an application-driven and human-centered perspective, investigating a concrete use case developed in collaboration with direct stakeholders~\citep{blilihamelin2025stoptreatingaginorthstar, rolnick2024applicationdriveninnovationmachinelearning}. Specifically, we focus on instructional support through AI-powered responses to student questions on a course portal. In large-enrollment courses (400+ students), such portals serve as a critical complement to in-person support; however, during peak periods, such as exams, the volume of questions can overwhelm the teaching staff. In our case study of a Calculus I course, nearly 700 questions were posted during three weeks of exam preparation, highlighting the need for scalable, reliable, and pedagogically aligned support.

To address this challenge, we design and evaluate a response-generation model tailored to the Calculus I course. Our aim was not simply to automate responses but to emulate the instructional style of course staff, producing accurate, contextually relevant answers that would feel familiar to students. We curated a dataset of a total of 2,738 student–instructor Q\&A pairs, augmented with lecture notes and exercises, and used this corpus to fine-tune a lightweight language model. To ensure a meaningful evaluation, we collaborated with experienced Calculus I instructors. Model responses were evaluated with respect to correctness, relevance, completeness, and alignment with question intent~\citep{ying2025benchmarkinghumanlikeintelligencemachines}. Following model validation, we deployed the model pipeline in the Fall 2025 and collected qualitative and quantitative feedback from both the course instructor and students.\footnote{This research was approved by the institutional IRB (HREC000626).} Our research process is outlined in Figure \ref{fig:layout_of_work}.

These are the research questions we aim to answer:
\begin{itemize}
    \item RQ1: How well can a course-specific generative AI model produce accurate, relevant, and pedagogically aligned responses to student questions in large-enrollment courses?
    \item RQ2: What types of student questions (e.g., easy, medium, hard) expose the strengths and weaknesses of AI-generated responses, and how do these limitations affect instructional support?
    \item RQ3: In what ways do the responses generated by the model differ in quality and perceived helpfulness from those of the instructors?
    \item RQ4: What dimensions of responses can be reliably assessed with automated evaluation?
    \item RQ5: How does the model perform during real-world deployment, as reflected by the instructor’s actions, such as endorsing, complementing, or deleting the model’s responses?
    \item RQ6: How do students perceive the model’s responses during course use?
\end{itemize}

Our findings underscore that the role of generative AI in education is not to replace instructors but to expand their capacity while maintaining opportunities for oversight and feedback. Although models sometimes produce fluent yet inaccurate responses or misinterpret student questions, structured integration with instructor review can mitigate these risks. Importantly, such oversight not only ensures instructional quality but also provides students with opportunities to critically reflect on AI-generated answers, helping them recognize both strengths and limitations of the technology. By situating our study in a large enrollment course with high instructional demand, we illustrate how human–AI collaboration can reduce staff workload while maintaining pedagogical rigor.

\section{Related work}

Research on intelligent tutoring systems (ITS) precedes the rise of large language models (LLMs), with early systems such as AutoTutor, Cognitive Tutor, and ALEKS exploring adaptive feedback, structured practice, and personalized learning support~\citep{Luckin, Holstein_McLaren_Aleven_2019, Guo2021ITS}. Complementary work on analytics dashboards emphasized classroom orchestration and teacher workload management~\citep{Verbert}, while learning sciences research highlighted the sociotechnical dimensions of scaling instructional support in large classes~\citep{Jerrim}. These strands established enduring concerns around scalability, orchestration, and teacher agency that continue to frame contemporary AI deployments in education.

Recent work investigates how LLMs can act as tutors, teaching assistants, or collaborators in learning contexts. Several benchmarks have been introduced to evaluate tutoring quality and responsiveness to learner needs, including MathTutorBench for alignment with student goals~\citep{macina2025mathtutorbenchbenchmarkmeasuringopenended}, BIPED for bilingual and adaptive instruction~\citep{kwon2024bipedpedagogicallyinformedtutoring}, and datasets embedding pedagogical strategies for domains such as reading comprehension~\citep{fateen2024developingtutoringdialogdataset}. Other benchmarks simulate authentic tutoring interactions, such as TutorEval~\citep{chevalier2024languagemodelssciencetutors} and TutorUp~\citep{pan2025tutorupstudentssimulatedtraining}. Beyond controlled datasets, real-world deployments underscore both potential and limitations: an AI math tutor in Ghana demonstrated affordability and effectiveness at scale~\citep{henkel2024effectivescalablemathsupport}, while other studies highlight persistent concerns over shallow pedagogical reasoning in LLM outputs~\citep{hicke-etal-2023-assessing}.

In parallel, human-computer interaction (HCI) and learning sciences research foreground human-centered and participatory approaches. Studies show that effective systems require stakeholder involvement throughout design and evaluation~\citep{Jurenka2024TowardsRD, saxena2025aimismatchesidentifyingpotential}. Teachers are increasingly positioned as co-designers of LLM-powered tools, with recent studies showing applications for grading, personalized feedback, project management, and curriculum planning~\citep{Ravi_2025}. Systems such as CT4ALL combine automated assessment with teacher input in computational thinking environments~\citep{Troiano}. AI agents, which enhance classroom collaboration, are often developed through iterative co-design with educators and learners~\citep{Doherty}. Teachable agents in computer science education demonstrate both promise and risks, where sustained motivation depends on careful integration with teaching practices~\citep{Rogers}.  Students’ perspectives add another layer of complexity. Research highlights both enthusiasm and caution: students appreciate AI’s availability but worry about over-reliance, reduced critical engagement, and diminished independence in reasoning. Recent frameworks argue that educational AI should empower learners, encourage critical thinking, and protect student agency rather than displacing genuine cognitive effort~\citep{favero2025aitutorsempowerenslave}. Broader policy and ethics perspectives emphasize the need for AI literacy, institutional support, and inclusive governance structures~\citep{fuligni2025wouldwantaitutor}.

To evaluate LLMs in the context of mathematical knowledge, researchers have developed a growing ecosystem of benchmarks that assess their reasoning capabilities. These include general-purpose math benchmarks (e.g., GSM8k~\citep{cobbe2021trainingverifierssolvemath}, AIME~\citep{dizhang2025}, MATH500~\citep{lightman2023lets}), as well as specialized resources for university-level calculus~\citep{chernyshev2025umathuniversitylevelbenchmarkevaluating}, Olympiad-level geometry~\citep{SolvingOlympiad53097}, and multi-step tool-augmented reasoning~\citep{zhang2023evaluatingimprovingtoolaugmentedcomputationintensive}. Instruction-tuning corpora such as OpenMathInstruct \citep{toshniwal2024openmathinstruct118millionmath} further extend math-specific capabilities. However, most mathematical benchmarks privilege correctness over other instructional dimensions, leaving questions of pedagogical alignment underexplored.

Our work shifts focus from synthetic benchmarks to authentic, messy student–instructor interactions in a large enrollment Calculus I course. Rather than evaluating correctness alone, we examine whether AI responses clarify concepts, provide guided hints, and align with instructors’ pedagogical values. In mathematics education, such alignment emphasizes rigor, logical progression, and scaffolding. Consistent with prior work stressing rigorous human evaluation and participatory design, we collaborated closely with the course instructor to shape model development and evaluation methodology. The resulting framework was then validated through human evaluation conducted with multiple instructors who regularly teach the same course.

\section{Task Framing and Dataset}
\subsection{Task Definition}
We frame the task as single-turn educational response generation, where a model produces subject-specific clarifications or explanations in response to a student's question. This design departs from traditional tutoring systems that rely on multi-turn dialogue and step-by-step derivations. Instead, our focus is on generating correct, concise, and context-aware responses that align with the instructional style of the course. Each input consists of a student question and the relevant course context (e.g., an excerpt from lecture notes or an exercise). The output is expected to be pedagogically appropriate and provide enough explanation to support understanding without overwhelming students with extraneous details~\citep{Hazan-Liran-2024, Cognitive}. 


\subsection{Data Collection}
\begin{figure*}[t]
      \centering
      \includegraphics[width=\textwidth]{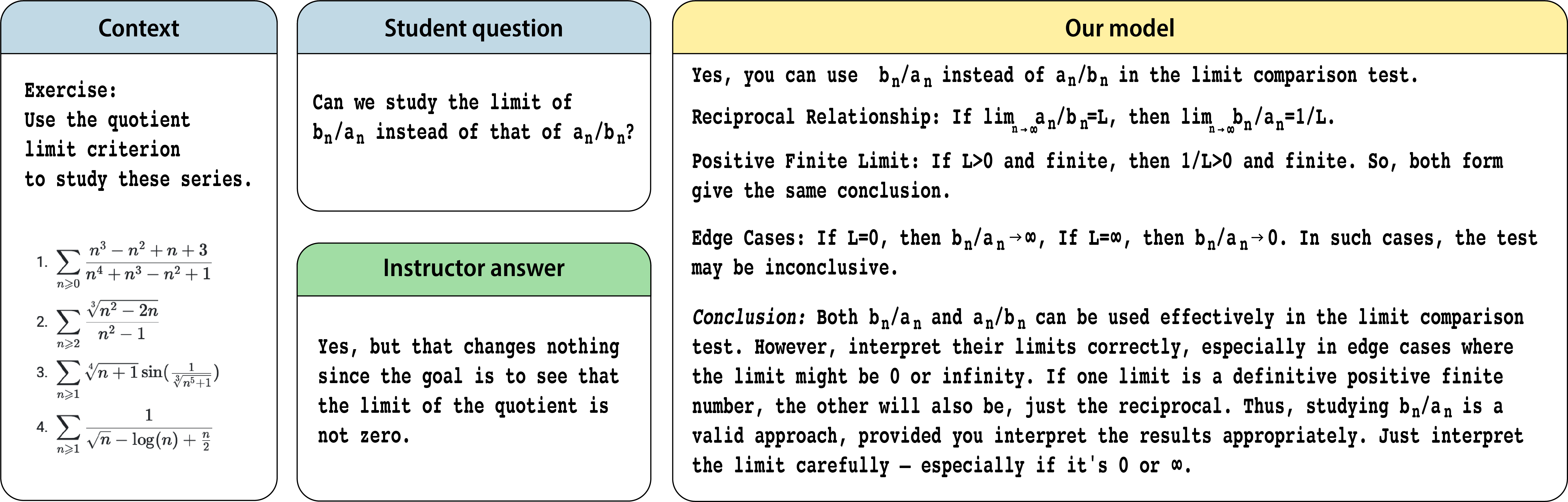}
      \caption{Example from the dataset, showing the context, student question, instructor answer, and our model response.}
      \Description{}
      \label{fig:main_example}
\end{figure*}

Our dataset is based on a university-level Calculus I course. The course portal, designed and maintained by the instructor, integrates interactive lecture notes with an embedded Q\&A forum. Each paragraph or exercise is directly linked to a dedicated discussion space, where students can pose questions that remain visible to peers. This design encourages engagement and provides rich contextual grounding for both questions and answers. An example of our dataset with generated response is shown in Figure~\ref{fig:main_example}. 

During the two semesters the course platform was in operation, students submitted more than 3,000 questions. The responses were provided directly by the instructor or drafted by teaching assistants and subsequently verified by the instructor, ensuring high reliability of the answers.

\subsection{Dataset Curation and Processing}
To build the corpus, we removed low-content questions (e.g., lecture logistics) and answers consisting solely of links to lecture notes. To eliminate the effect of language on model performance, we translated all data from French into English using GPT-4o~\citep{openai2024gpt4ocard}. The resulting dataset includes 2,738 question–answer pairs, each paired with a contextual segment averaging roughly a paragraph in length. In total, 939 lecture note files yielded 724 segments used as grounding context. Full dataset statistics are reported in Appendix~\ref{app:datastat}.


In our dataset, we have a broad range of questions posed by students.  Some are similar to those in the existing benchmarks, like TutorBench~\citep{macina2025mathtutorbenchbenchmarkmeasuringopenended}, where students want some explanation: 

\newcommand{\StudentQuote}[1]{%
    \par\vspace{0.5em}\noindent
    \colorbox{gray!10}{%
        \begin{minipage}{\dimexpr\linewidth-2\fboxsep}
            \small\itshape %
            #1
        \end{minipage}%
    }%
    \par\vspace{0.5em}%
}

\StudentQuote{
``I don't understand why we have to use $P(n+1)$. Should we compare both sides of the inequality with this method?''
}

Some questions illustrate the challenges students face in articulating their intent, providing authentic examples of misunderstanding.

\StudentQuote{
``For part one, is it sufficient to say that since the series of $a_{n}$ converges absolutely, then all of its subsequences converge, and thus each of their general terms converges to 0. If we sum them, we indeed get 0.

I am a little (a lot) lost, what does it mean to calculate? simplify? find an expression? how do I return to the expression for the geometric series?''
}

Questions requesting clarification of course concepts rather than problem solving:

\StudentQuote{
``Why does $k$ take the values: 0, 1, 2?''
}

\StudentQuote{
``For this kind of question, should you calculate the limit normally or use the definition of the limit?''
}

Questions asking for hints rather than full solutions:

\StudentQuote{
``I am unable to find a known series for exercise 4. The powers of 2 in the denominator that change with $n$ are causing me trouble. Would you have a hint? Thank you.''
}

Questions seeking targeted help with difficult or lengthy computations:

\StudentQuote{
``For question one, I don't understand the solution. Where did the $1/2$ come from?''
}












\subsection{Dataset Splits}

We split the 2,738 examples into 2,348 training, 240 validation, and 150 test examples. The test set was manually curated by the course instructor (20+ years of teaching experience) to reflect representative student questions. 

The course instructor manually categorized questions by difficulty:

\begin{itemize}
    \item Easy – Direct questions answerable in one to two sentences.
    \item Medium – Explanations requiring multiple core concepts (most common type).
    \item Hard – Deep conceptual questions that often extend beyond the given context.
\end{itemize}

\subsection{Data Augmentation}

Instructor answers were often concise, reflecting practical constraints of responding to hundreds of student queries. To prevent the model from learning overly brief response styles, we adopted an augmentation strategy inspired by OpenMathInstruct~\citep{toshniwal2024openmathinstruct118millionmath}. Specifically, we inserted intermediate thinking steps between questions and answers, providing richer scaffolding for model training. Two augmented versions were generated per question using DeepSeek-R1-Distill-Qwen-32B~\citep{deepseekai2025deepseekr1incentivizingreasoningcapability}, and a random subset of 200 was manually verified. This yielded a final dataset of 5,176 Q\&A pairs for fine-tuning. Further details are provided in Appendix~\ref{app:augmentation}.

\begin{figure*}[t]
  \centering
  \includegraphics[width=\textwidth]{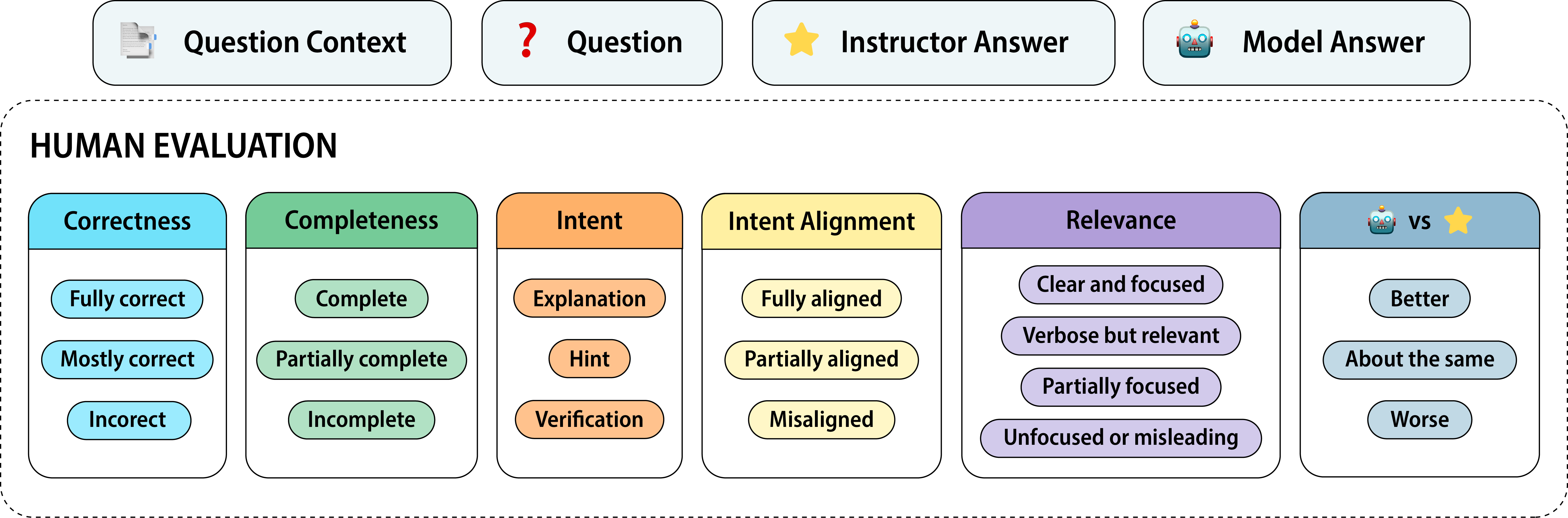}
  \caption{Survey layout showing the information presented to instructors (student question, context, instructor answer, and model response) and the evaluation criteria used for annotation.}
  \Description{}
  \label{fig:layout}
\end{figure*}

\section{Evaluation Design and Human Judgment}
Model development was conducted in close collaboration with the course instructor and guided by a human-centered evaluation approach. We performed two complementary evaluations: (1) a development-phase evaluation conducted by subject experts and (2) a post-deployment evaluation based on the instructor’s ongoing feedback and a student survey. This section describes the design and implementation of human-centered evaluation procedures, including the methods and instruments used in both stages.


\subsection{Instructor Evaluation Procedure. Development Phase.}\label{sec:human_eval}

For model development, we used a two-stage evaluation process outlined in Figure \ref{fig:layout_of_work}. In the first stage, the course instructor assessed responses to 40 representative student questions generated by several base models. Each response was rated for \textbf{Correctness}, \textbf{Relevance}, and \textbf{Completeness}. These ratings informed which models to fine-tune, and the same instructor then re-evaluated the fine-tuned versions to identify the most promising candidate.

In the second stage, we conducted a larger-scale evaluation of the best model. A panel of five experts (the course instructor plus four additional Calculus I instructors) reviewed 150 representative Q\&A pairs. To support reliability, all annotators rated a shared subset of 50 questions, and 25 questions were double-annotated. This design allowed us to balance the breadth of coverage with consistency between evaluators.

\subsubsection{Survey Instrument}
The survey was designed in consultation with the course instructor to highlight the pedagogical qualities of the responses. Models were always evaluated based on these criteria:
\begin{itemize}
    \item \textbf{Correctness} (3-point scale): \textit{Fully correct}, \textit{Mostly correct (minor issues)}, or \textit{Incorrect}. Only responses free from conceptual mathematical errors were considered fully correct. Minor issues (e.g., minor arithmetic slips) were tolerated if they did not impact understanding.
    \item \textbf{Completeness} (3-point scale): \textit{Complete}, \textit{Partially complete (minor omission)}, or \textit{Incomplete}.
    \item \textbf{Relevance} (4-point scale): This dimension assesses whether the response stays on topic without introducing distracting or irrelevant information.
    \begin{itemize}
        \item  \textit{Clear and focused} - directly answers the questions with appropriate details
        \item \textit{Verbose but relevant} - extra information is included, but still relevant to the topic
        \item \textit{Partially focused} - contains some tangential or loosely related information
        \item \textit{Unfocused/misleading}
    \end{itemize}    
\end{itemize}

In addition, the final 150-question evaluation with 5 experts included:

\begin{itemize}
    \item \textbf{Student Intent Classification}: Annotators identified the likely intent behind the student’s question using one of the following categories: \textit{Explanation}, \textit{Hint}, \textit{Verification}, \textit{Correction}, \textit{Request for full solution}, \textit{Clarification}, or \textit{Other/Unclear}.

    \item \textbf{Response Alignment}: Annotators judged how well the model’s response matched the identified intent:
    \begin{itemize}
        \item \textit{Fully aligned} – Response type clearly matches the student's request.
        \item \textit{Partially aligned} – Addresses the topic but in a different format than requested.
        \item \textit{Misaligned} – Misunderstands or ignores the student’s intent.
    \end{itemize}
    
    \item \textbf{Comparison to Instructor Answer}: The annotators compared the model response with the original answer: \textit{Better}, \textit{About the same}, or \textit{Worse}.
\end{itemize}

Annotators were shown the question context (lecture excerpt or exercise), the student’s question, the instructor’s answer, and the model’s answer (with intermediate reasoning steps). Figure~\ref{fig:layout} outlines the survey layout. The full survey is presented in Figures ~\ref{app:survey_1} and ~\ref{app:survey_2}.

\subsubsection{Annotator Expertise}
All annotators have substantial experience teaching Calculus I, with an average of 9.6 years of instructional experience. The primary course instructor, who collaborated with us throughout the evaluation design process, has more than 20 years of teaching experience. Because the course is taught in French, all annotators are fluent in both French and English to ensure accurate interpretation of student questions and model responses. To protect participant privacy, no additional demographic information was collected. Participation was voluntary, and informed consent was obtained in accordance with institutional review protocols (Appendix~\ref{app:Human_ann}).

\begin{figure*}[ht!]
 \centering
 \begin{subfigure}[t]{.32\textwidth}
     \centering
     \includegraphics[width=\textwidth]{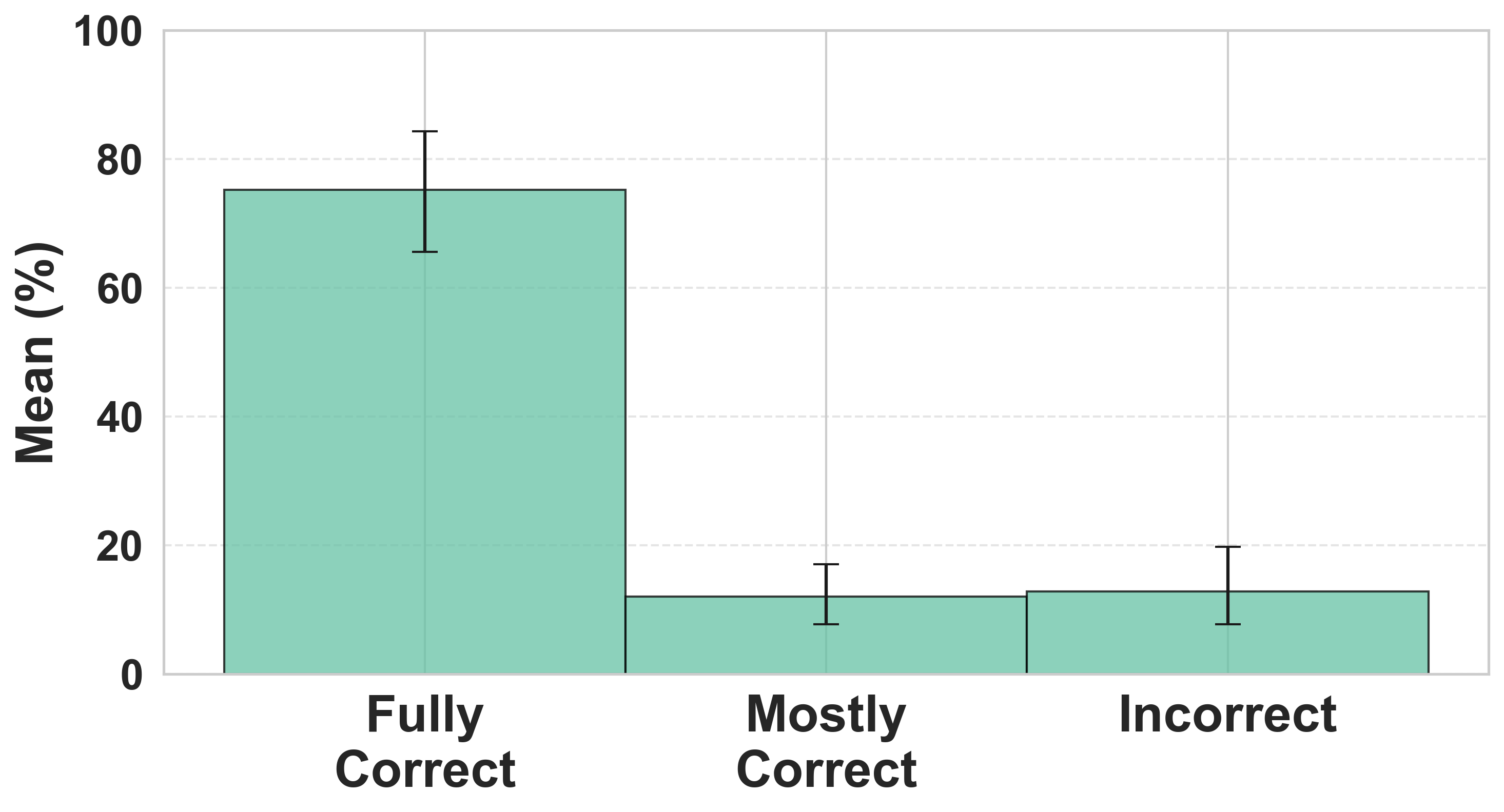}
     \caption{Correctness}\label{fig:corr_h}
 \end{subfigure}
 \begin{subfigure}[t]{.32\textwidth}
     \centering
     \includegraphics[width=\textwidth]{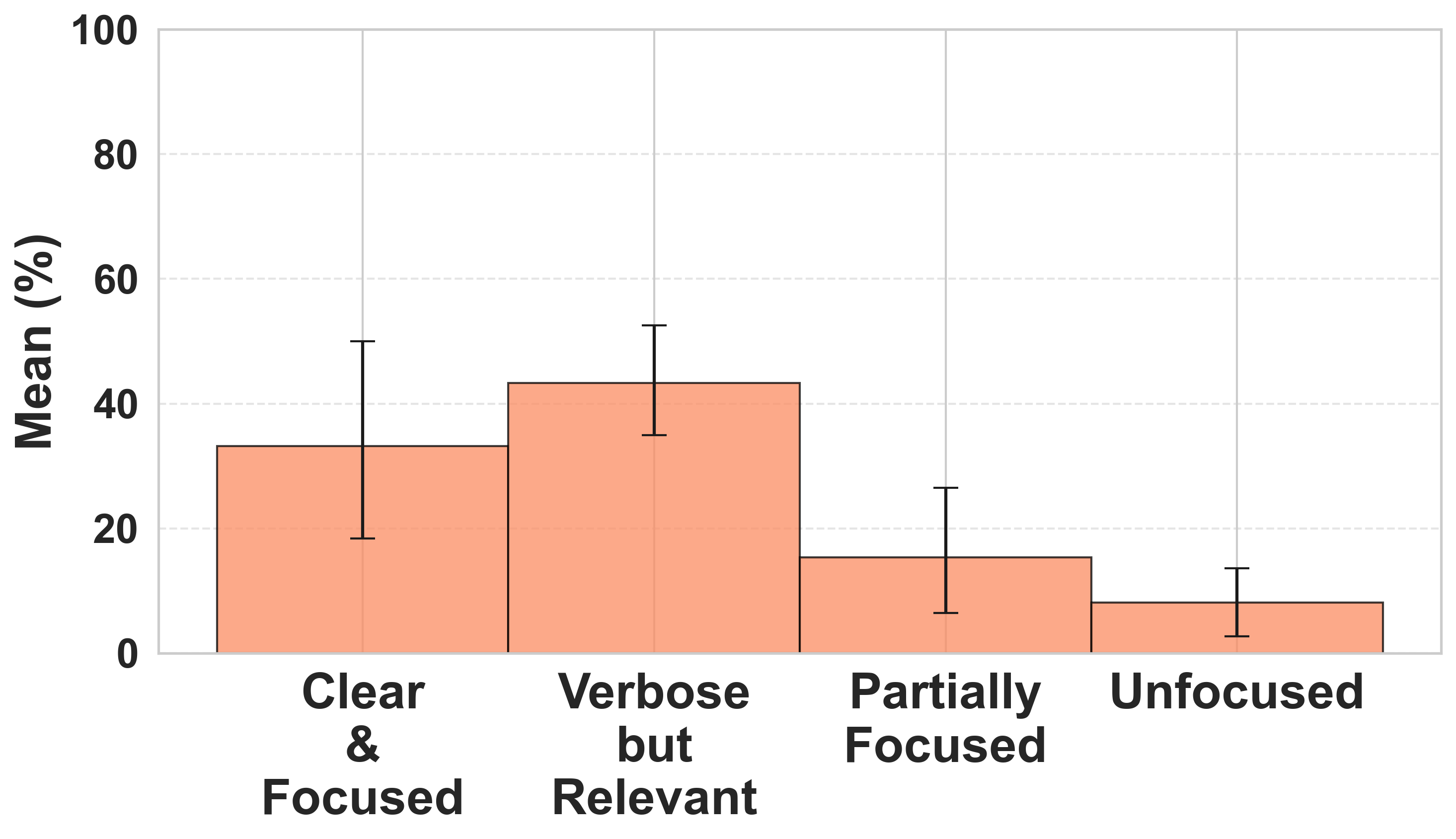}
     \caption{Relevance}\label{fig:rel_h}
 \end{subfigure}
 \begin{subfigure}[t]{.32\textwidth}
     \centering
     \includegraphics[width=\textwidth]{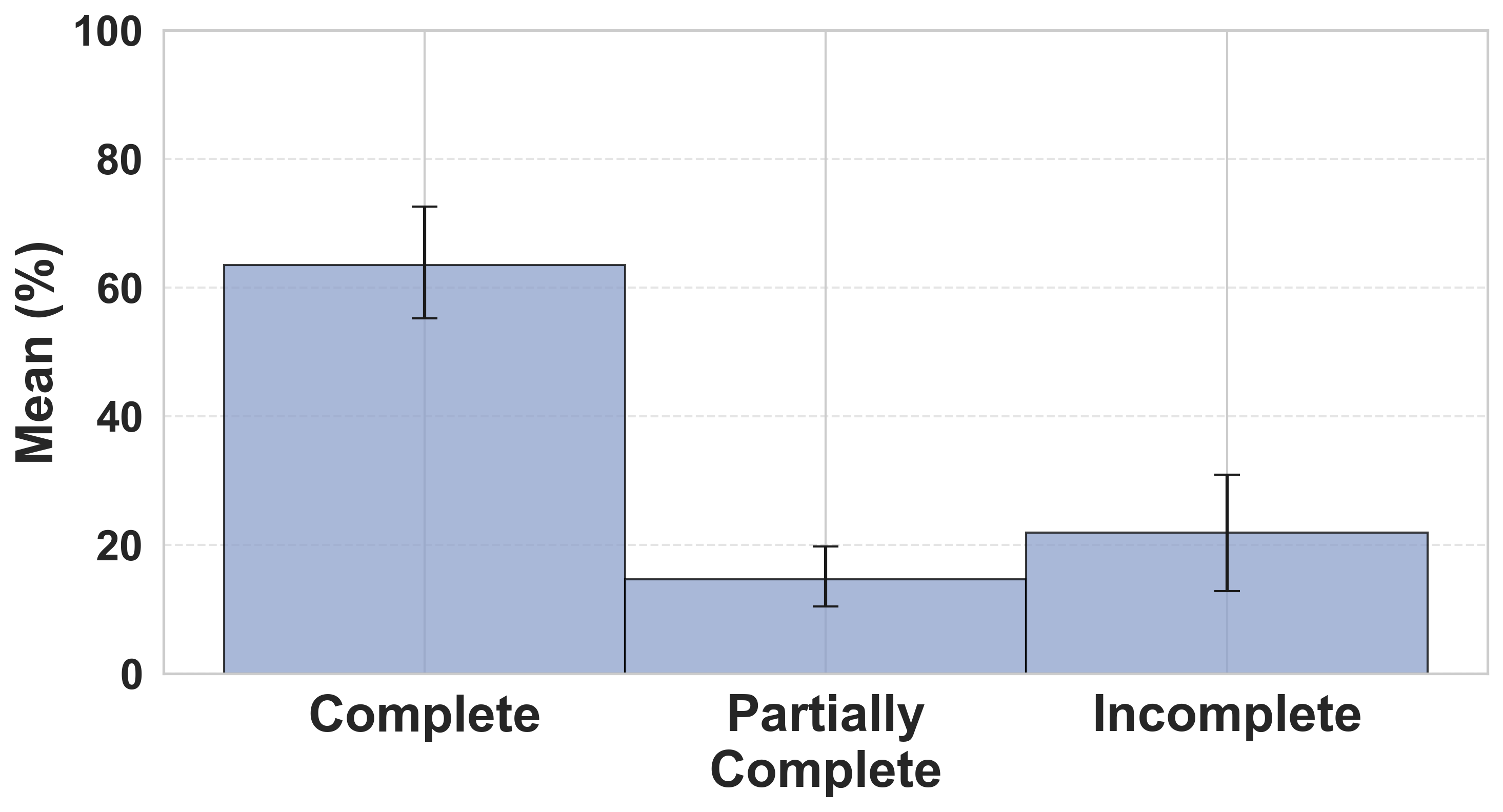}
     \caption{Completeness}\label{fig:comp_h}
 \end{subfigure}
 \caption{Human annotation results for the dataset of 150 questions on (a) Correctness, (b) Relevance, and (c) Completeness. 95\% confidence intervals are computed via non-parametric bootstrap with 10000 resamples. }
 \Description{}
 \label{fig:human_results}
\end{figure*}

\subsection{Automated Evaluation}

Building on promising work using LLMs for automated evaluation ~\citep{chiang2023large, Borges2024, xie2024gradelikehumanrethinking, mok2024usingailargelanguage}, we use GPT-o4-mini ~\citep{openai2025o3o4mini} to assess our best model’s responses on 150 questions and compare them with instructor ratings. This shows to what extent automated annotation can substitute for human instructor annotation.

\begin{figure*}[ht!]
 \centering

 \begin{subfigure}[t]{\textwidth}
     \centering
     \includegraphics[width=1.0\textwidth]{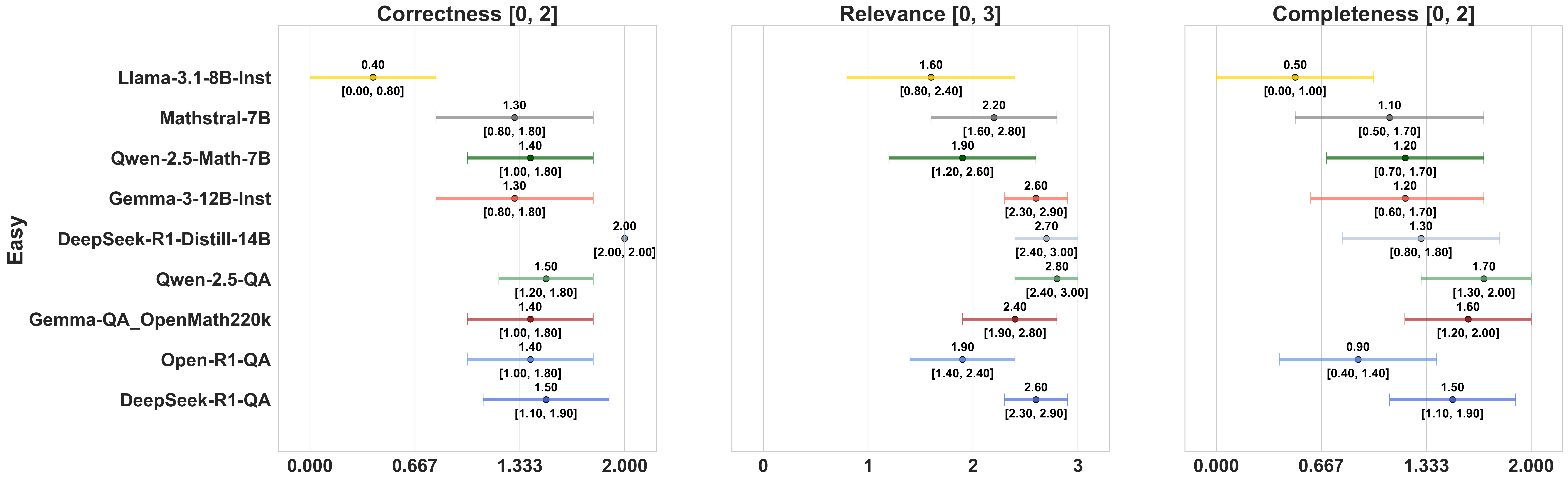}
     \caption{}\label{fig:easy}
 \end{subfigure}

 \begin{subfigure}[t]{\textwidth}
     \centering
     \includegraphics[width=1.0\textwidth]{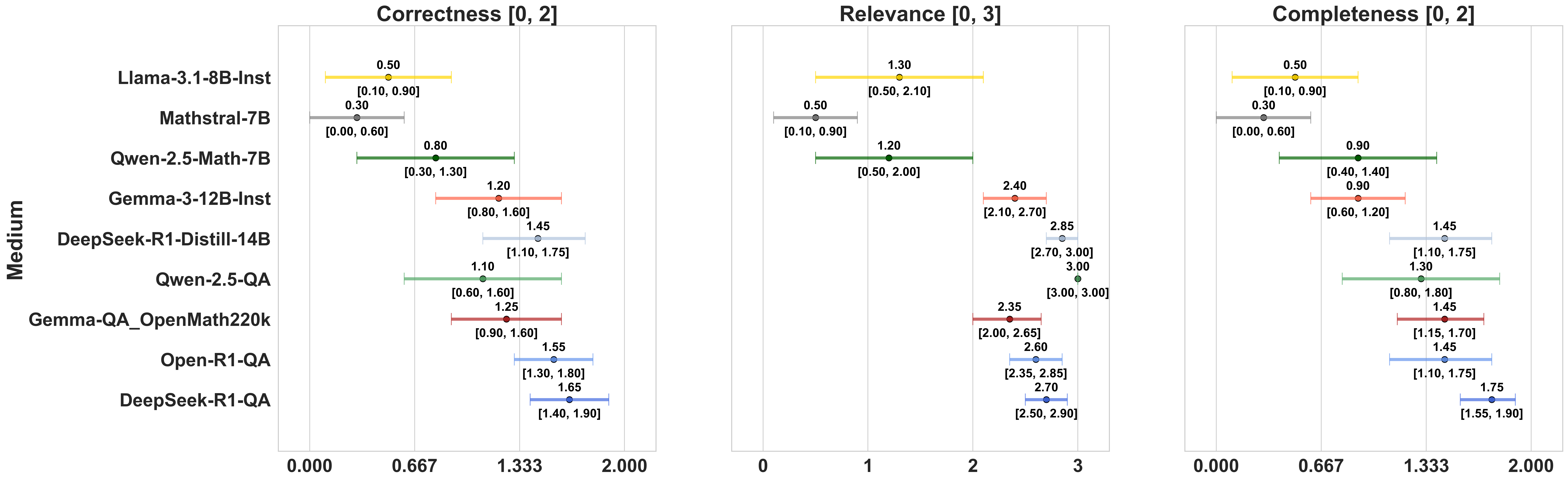}
     \caption{}\label{fig:medium2}
 \end{subfigure}

 \begin{subfigure}[t]{\textwidth}
     \centering
     \includegraphics[width=1.0\textwidth]{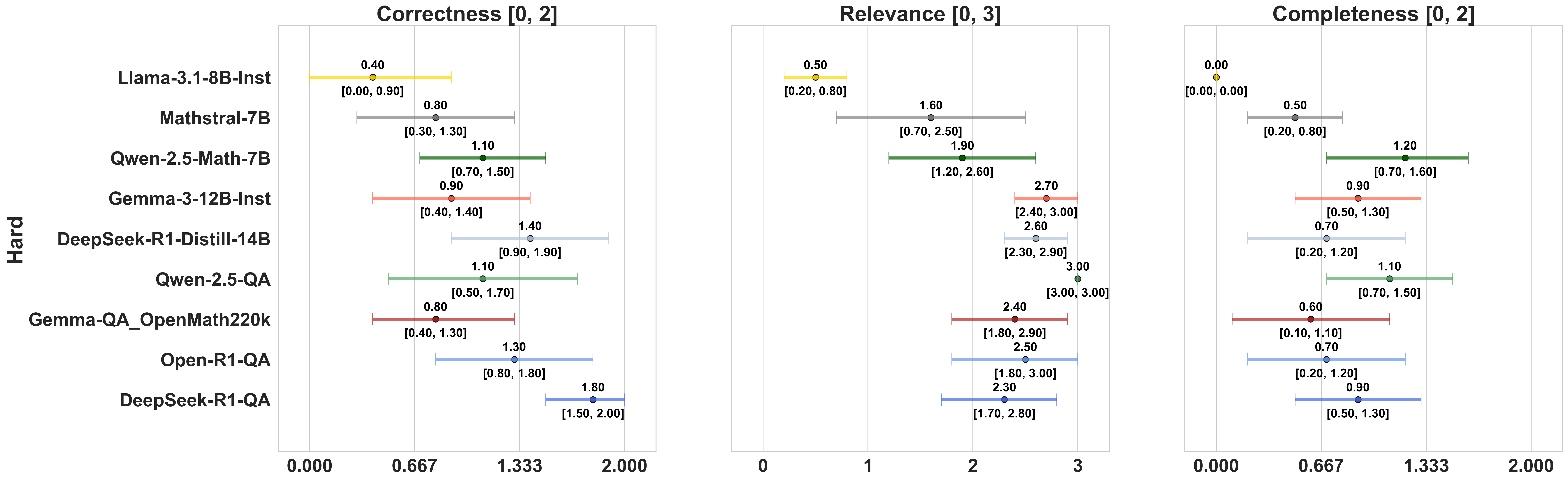}
     \caption{}\label{fig:hard}
 \end{subfigure}

 \caption{Evaluation of base and finetuned models on 40 test questions by a single expert. The results are shown by the question's difficulty: we have 10 examples of easy questions, 20 medium, and 10 hard ones. Base models: Llama 3.1, Mathstral, Qwen 2.5, Gemma 3, and DeepSeek R1 Distill. Models with ``-QA'' are finetuned on our dataset; ``OpenMath220k'' are finetuned on the respective math dataset. Scores shown are averages. 95\% confidence intervals are computed via non-parametric bootstrap with 10000 resamples.}
 \Description{}
 \label{fig:model_difficulty_eval}
\end{figure*}

\subsection{Student Evaluation Procedure}\label{sec:student_eval}

The student survey assessed learners’ satisfaction with the automated responses, their trust in the system, and their perceptions of the AI use in the course. The survey was administered in French, took approximately four minutes to complete, and participation was voluntary. The full survey is presented in Appendix \ref{app:student_survey}.

\noindent \textbf{Background information.} We first collect contextual data on students’ (1) frequency of posting questions on the question forum; (2) typical types of questions they ask. The categories for this item were proposed by the course instructor based on observed common questions over the years of teaching, with an option for students to add their own category; and (3) self-reported familiarity with the course material, rated on a five-point scale from Very Poor (“I do not understand basic concepts”) to Very Good (“I understand the material thoroughly and can explain it to others”).

\noindent \textbf{Perceptions of model performance.} Students rated the extent to which the model’s responses aligned with their intended question on a four-point Likert scale. For instance, if they ask a model for a hint, does the model respond with a hint?

\noindent \textbf{Trust and perceived qualities.} Students evaluated their trust in the model responses, the degree to which the model’s explanations resembled the instructor’s style, and the usefulness of the model’s responses, each on a five-point Likert scale from Strongly disagree to Strongly agree.

\noindent \textbf{Perceived impact.} To assess broader attitudes toward deployment, students rated their agreement (five-point Likert scale) with statements regarding the model’s impact: whether it saved them time compared to waiting for instructor responses and whether they would like similar models in other courses.

\noindent \textbf{Open-ended questions.} Finally, the survey included free-response questions asking students to elaborate on (1) what they appreciated about the model’s responses, (2) what aspects they felt could be improved, and (3) what factors increased or decreased their trust in the model’s answers.

\subsection{Instructor Post-Deployment Evaluation}
Following deployment, we implemented an oversight procedure in which the instructor reviews each model-generated response within 24 hours of a student posting a question. For every response, the instructor selects one of four actions: (1) endorse the response as fully acceptable; (2) endorse with edits, revising the model’s answer; (3) add a supplemental instructor comment while keeping the model’s original response visible; or (4) delete the model’s response if it was incorrect, unclear, or otherwise inappropriate. 
This review process serves both as a quality-control mechanism and as a means of assessing the model’s pedagogical reliability during authentic use. In addition, throughout the Fall 2025 semester, we logged quantitative indicators of system use, including the total number of student questions submitted, the number of unique student users, and the number of ``likes'' students gave to model responses they found helpful.

\section{Instructor-Guided Model Development}\label{model}

We prioritize small-to-medium models (7–14B parameters) to balance performance with accessibility in educational contexts. These models are affordable for institutions with limited resources, easier to host locally for privacy, and more practical to integrate into classroom workflows. We considered several recent open-source models with strengths in mathematical reasoning or instructional alignment (e.g., Llama 3.1\citep{grattafiori2024llama3herdmodels}, Qwen 2.5\citep{qwen2025qwen25technicalreport}, Mathstral\citep{jiang2023mistral7b}, Open R1\citep{openr1}, DeepSeek R1\citep{deepseekai2025deepseekr1incentivizingreasoningcapability}, and Gemma 3\citep{gemmateam2025gemma3technicalreport}). 

\textbf{Model Selection for Fine-Tuning:} To ground model choice in pedagogical quality rather than purely quantitative metrics, the course instructor evaluated each base model on 40 representative student questions. This required approximately 3.5 hours per model and revealed that DeepSeek R1 performed best among base models. We also fine-tuned Qwen 2.5, Gemma 3, and Open R1 as promising models.

\textbf{Fine-Tuning Strategy:} We applied two complementary fine-tuning approaches. Math-specialized models (e.g., DeepSeek R1 and Qwen 2.5) were directly fine-tuned on course-specific Q\&A pairs, augmented with reasoning steps. General-purpose models (e.g., Gemma) were first adapted on a calculus-focused pretraining set (OpenR1-Math-220k~\citep{openr1_math_220k}), then fine-tuned for alignment with the course. To guard against catastrophic forgetting, models were also checked against standard mathematical benchmarks (AIME 24, AIME 25, MATH 500, GPQA:Diamond~\citep{rein2023gpqagraduatelevelgoogleproofqa, LLMReasoningBenchmark}, and UMATH~\citep{chernyshev2025umathuniversitylevelbenchmarkevaluating}), though our emphasis remained on instructor-led evaluation. Full details of fine-tuning configurations are provided in Appendix~\ref{app:model}.

\textbf{Final Model Selection:} The course instructor evaluated the fine-tuned models on the same set of 40 test questions. Figure~\ref{fig:model_difficulty_eval} shows the results for both base models and the fine-tuned ones. All models fine-tuned on our data produced more consistent and pedagogically aligned responses. Our fine-tuned version of DeepSeek R1 performed particularly well on medium- and hard-difficulty questions, and was selected for in-depth evaluation by a panel of five expert instructors (Section~\ref{sec:human_eval}). The benchmark results (Table~\ref{tab:math-results}) confirmed strong general reasoning ability, but our final decision was guided primarily by expert judgments of response quality in authentic educational contexts. Cost-wise, our system is approximately 56× cheaper than GPT-5 or Gemini 2.5 Pro, and 3.5× cheaper than DeepSeek. More details on the final cost of the pipeline are in Appendix~\ref{app:costs}. 

\textbf{Model Improvement Based on Instructor's Feedback After Deployment:} After the final evaluation by five instructors, we deployed the model on the course portal for the Fall 2025 semester. During the semester, we met weekly with the course instructor to review model performance and identify issues in real use. Based on this feedback, we introduced the following model pipeline:
\begin{itemize}
    \item Image-to-text conversion: Many students submitted questions as photos of their notes, so we added an OCR step to convert images into text for processing.
    \item Question-type classification and specialized prompts: Using an instructor-developed taxonomy, we added automatic classification of student questions and created tailored prompts for each type. The instructor could adjust prompts as needed. The categories included: lecture-content questions, misunderstandings of exercises, requests for hints or starting guidance, answer verification, full-solution requests, unrelated questions, misplaced course questions, and others. The model details used for this task are provided in Appendix ~\ref{app:adjustements}.
\end{itemize}

These adjustments were introduced incrementally during the semester, reflecting the need for flexible, instructor-informed refinement as real usage and student demand surfaced new requirements. As small language models continue to improve, future iterations may incorporate stronger architectures to enhance pedagogical alignment further. The model, data for finetuning, benchmark, and code are available in this \href{https://github.com/botafogo-EPFL/CaLlm_pub}{GitHub repository}.


\section{Evaluation Results}
We begin by presenting the human evaluation of the final model on a benchmark of 150 Q\&As and compare these results with automated evaluation. This analysis addresses the model’s effectiveness, variation across question types, comparison to instructor responses, and alignment with automated evaluation (RQ1–RQ4). We then examine the post-deployment performance of our model pipeline based on the instructor’s actions during live use (RQ5) and student perceptions (RQ6).

\subsection{RQ1: Effectiveness of Course-Specific AI Responses}
Overall, the model produced fully correct answers in 75.3\% of cases, with an additional 12\% rated as mostly correct (Figure~\ref{fig:human_results}). Responses were complete in 64\% of cases and partially complete in 14.7\%. Regarding relevance, 33.3\% of responses were clear and concise, while 43.3\% were verbose yet still on topic. Correctness received the most consistent ratings across annotators, reflected in narrow confidence intervals.

Inter-annotator agreement averaged 83.2\% overall, with correctness achieving the highest consensus (92\%) and relevance the lowest (62\%), reflecting its more subjective interpretation. Annotators reported that ``verbosity'' could sometimes be seen as supportive elaboration or as unnecessary detail, depending on the question context. Detailed agreement metrics are provided in Table~\ref{tab:agreement}.

\begin{table}[t]
\centering
\small
\caption{Agreement (\%) across annotators by metric and difficulty level.}
\begin{tabular}{l l r}
\toprule
Metric & Difficulty & Agreement (\%) \\
\midrule
\multirow{4}{*}{Correctness} 
       & All        & 92.00 \\
       & Easy       & 93.33 \\
       & Hard       & 86.67 \\
       & Medium     & 95.00 \\
\midrule
\multirow{4}{*}{Relevance} 
       & All        & 62.00 \\
       & Easy       & 53.33 \\
       & Hard       & 66.67 \\
       & Medium     & 65.00 \\
\midrule
\multirow{4}{*}{Completeness} 
       & All        & 86.00 \\
       & Easy       & 80.00 \\
       & Hard       & 93.33 \\
       & Medium     & 85.00 \\
\midrule
\multirow{4}{*}{Expectation alignment} 
       & All        & 90.00 \\
       & Easy       & 93.33 \\
       & Hard       & 86.67 \\
       & Medium     & 90.00 \\
\midrule
\multirow{4}{*}{Comparison} 
       & All        & 86.00 \\
       & Easy       & 93.33 \\
       & Hard       & 73.33 \\
       & Medium     & 90.00 \\
\bottomrule
\end{tabular}

\label{tab:agreement}
\end{table}

\begin{figure*}[t]
  \centering
  \includegraphics[width=0.85\textwidth]{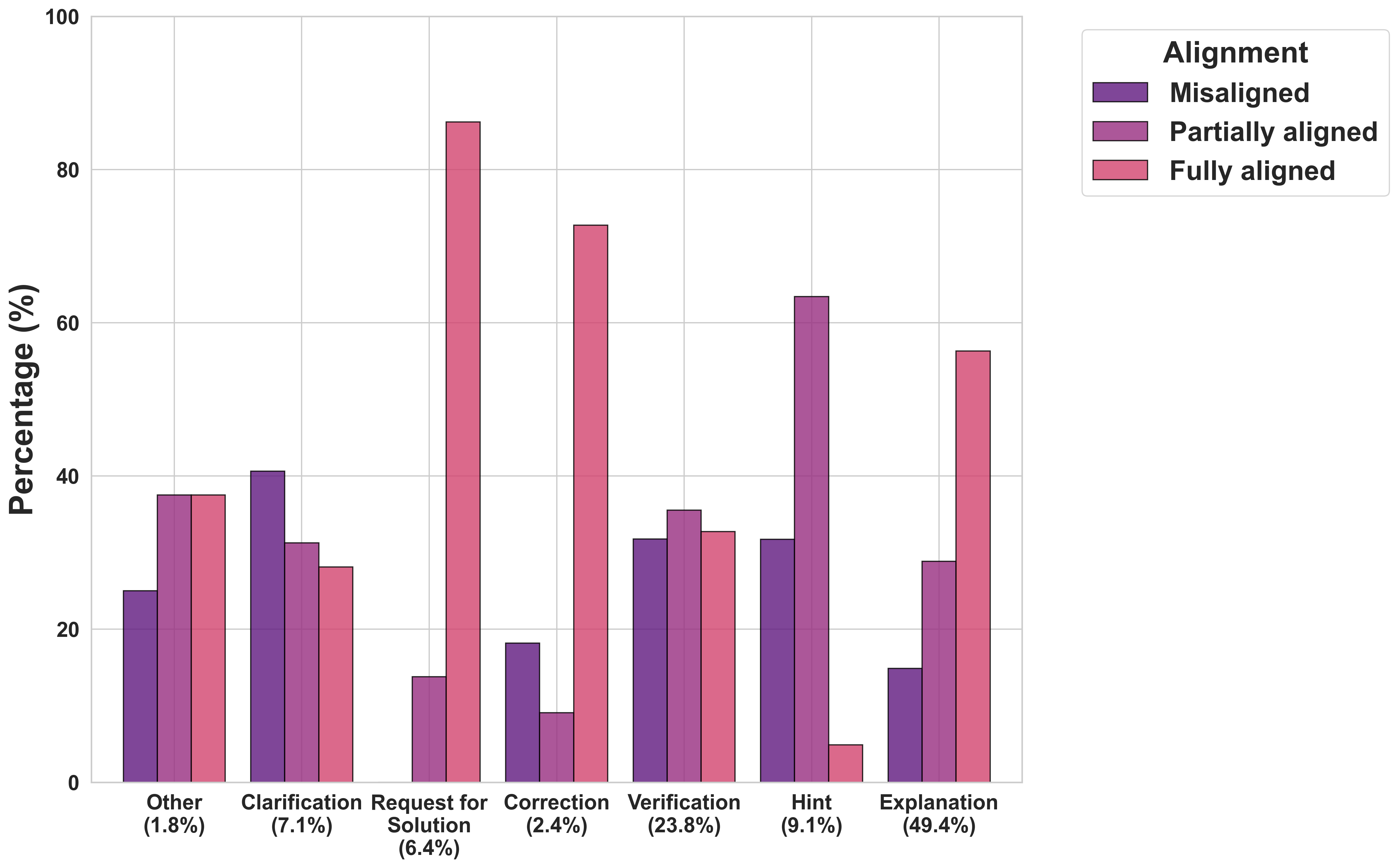}
  \caption{Distribution of model response alignment across question types for a set of 150 Calculus I questions. Each question type (e.g., Explanation, Hint, Clarification) is labeled with its proportion in the dataset. Bar colors represent the degree of alignment between the model's response and the student question intent: \textit{Misaligned}, \textit{Partially aligned}, or \textit{Fully aligned}.}
  \Description{}
  \label{fig:aliognment}
\end{figure*}

We also examined how well the model aligned with the inferred intent of each student question. The most frequent intent was Explanation (49\%), followed by Verification (24\%) and Hints (9\%). The model aligned best with explanation, correction, and solution requests, but struggled with hints, verifications, and clarifications (Figure~\ref{fig:aliognment}). Across the dataset, the model achieved full intent alignment in 44.7\% of cases.

Finally, instructors compared model-generated responses with gold answers provided by the course instructor. In 36\% of cases, the model’s response was rated as equally good as or better than the instructor’s. Most of these occurred in explanation-type questions (40.5\%), which instructors noted are also the most common form of student inquiry. They emphasized that in these cases, the model’s detailed step-by-step reasoning could be particularly helpful for students who require elaboration beyond a minimal reply.

These results show us that a lightweight, course-tailored model can reach a level of accuracy that is pedagogically useful without requiring massive architectures. However, variability in perceived relevance points to the need for interaction designs that allow instructors or students to adjust response style (e.g., concise vs. detailed).


\subsection{RQ2: Strengths and Weaknesses Across Question Difficulty}
Performance varied systematically with question difficulty. The model performed strongest on medium-difficulty questions, the most common category, where responses were more accurate, complete, and contextually appropriate (Figure~\ref{fig:mean_per_dimension_with_ci}). For easy questions, instructors observed a tendency toward over-explanation, which, while accurate, may not always be aligned with student expectations. For hard questions, the model often identified the relevant topic but failed to provide sufficient depth, leading to partial or incomplete answers.

Qualitative feedback from instructors suggests that medium-difficulty questions represent a ``sweet spot'' where the model’s elaborative style is most helpful. In contrast, addressing complex questions still requires deep pedagogical reasoning and diagnostic skill that remains the domain of experienced instructors.

Thus, we find that AI assistance is most effective in relieving routine instructional load (e.g., medium-difficulty explanation requests), while still requiring human expertise for advanced or ambiguous cases. A hybrid workflow, where the system addresses common queries and flags challenging ones for instructor review, may maximize both scalability and instructional quality.

\subsection{RQ3: Comparing Model and Instructor Responses}

We complemented quantitative evaluation with qualitative feedback from post-annotation interviews with instructors.

Annotators noted that the fine-tuned model closely mirrored the conventions of the course: it adopted the instructor’s notation, referenced familiar examples, and produced responses that felt consistent with classroom practice. This alignment was seen as beneficial for students, reducing cognitive overhead and fostering continuity between in-class instruction and online support.

Although the model often produced answers that were correct, complete, and appropriately detailed, it was still rated lower than instructor responses in 64\% of cases. Instructors explained that this difference reflects not only correctness but also the pedagogical nuance embedded in human answers. Whereas instructors often tailor their responses to the perceived needs of the student, like providing a quick confirmation, a targeted hint, or a carefully crafted clarification, the model’s default strategy was to generate extended explanations. This elaboration was perceived as valuable for students who benefit from additional support, but potentially excessive for advanced learners seeking concise guidance.

Instructors highlighted several key use cases where this distinction matters. For explanatory questions, the model’s detail was an asset, sometimes even surpassing instructor answers. For verification, hinting, and clarification tasks, however, the model struggled to provide responses that were appropriately scoped. These findings suggest that while elaboration can be supportive, there is no one-size-fits-all response style: different students benefit from different levels of detail.

A recurring concern was that when errors did occur, they were typically subtle rather than obvious. Such errors may be difficult for students to detect when working independently, raising risks if the system is used without oversight. Instructors emphasized that embedding the model into a hybrid workflow, where human staff can endorse, refine, or correct responses, could balance the strengths of both parties: scalable elaboration from the model, combined with pedagogical precision and contextual judgment from instructors.

These results highlight the importance of designing AI systems that allow for adaptive response styles and instructor mediation. Rather than assuming a uniform notion of ``helpfulness,'' systems should enable responses that can be tuned to student needs and embedded within instructional workflows that provide oversight and error correction.

\begin{figure*}[ht!]
 \centering
 \begin{subfigure}[t]{.32\textwidth}
     \centering
     \includegraphics[width=\textwidth]{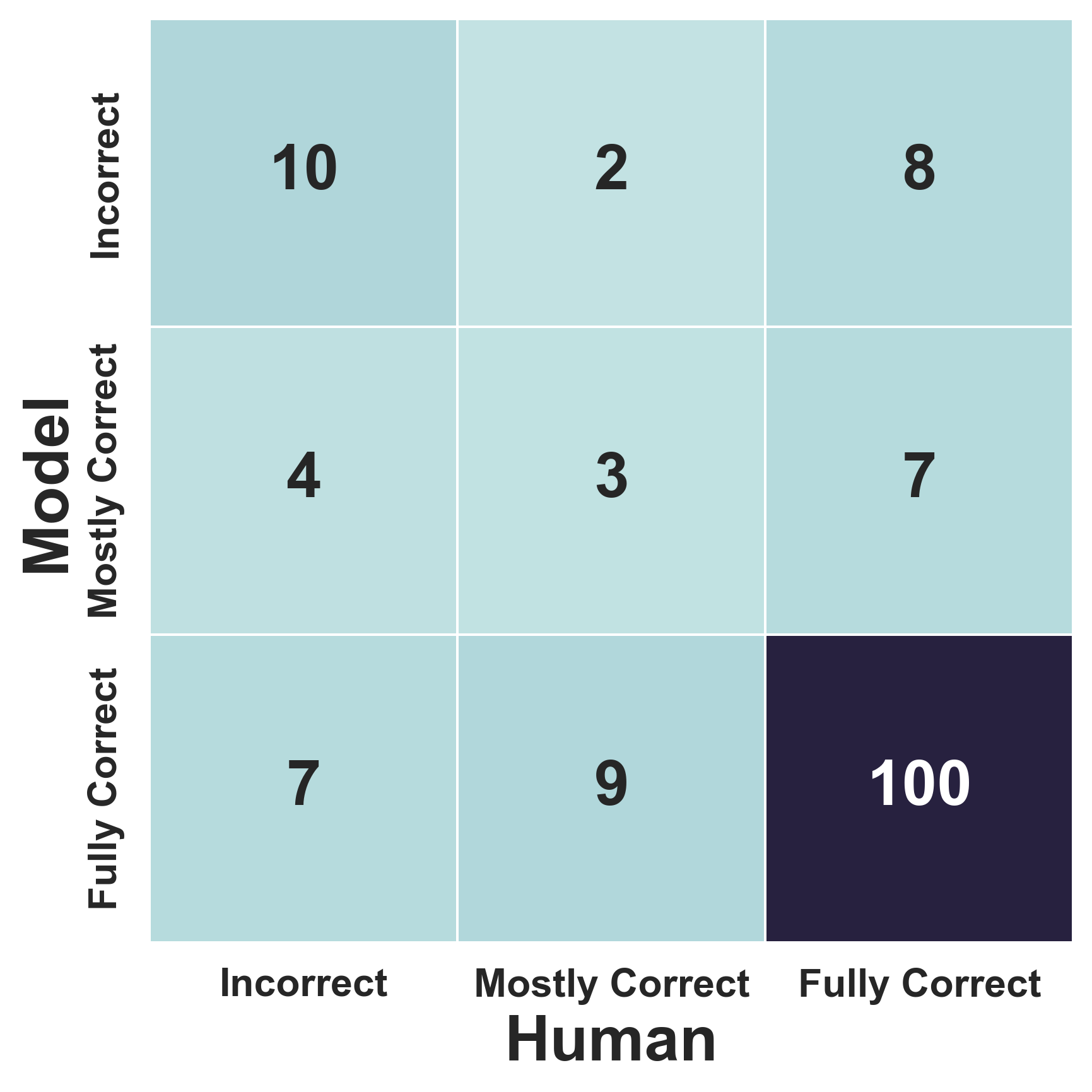}
     \caption{Correctness}\label{fig:corr}
 \end{subfigure}
 \begin{subfigure}[t]{.32\textwidth}
     \centering
     \includegraphics[width=\textwidth]{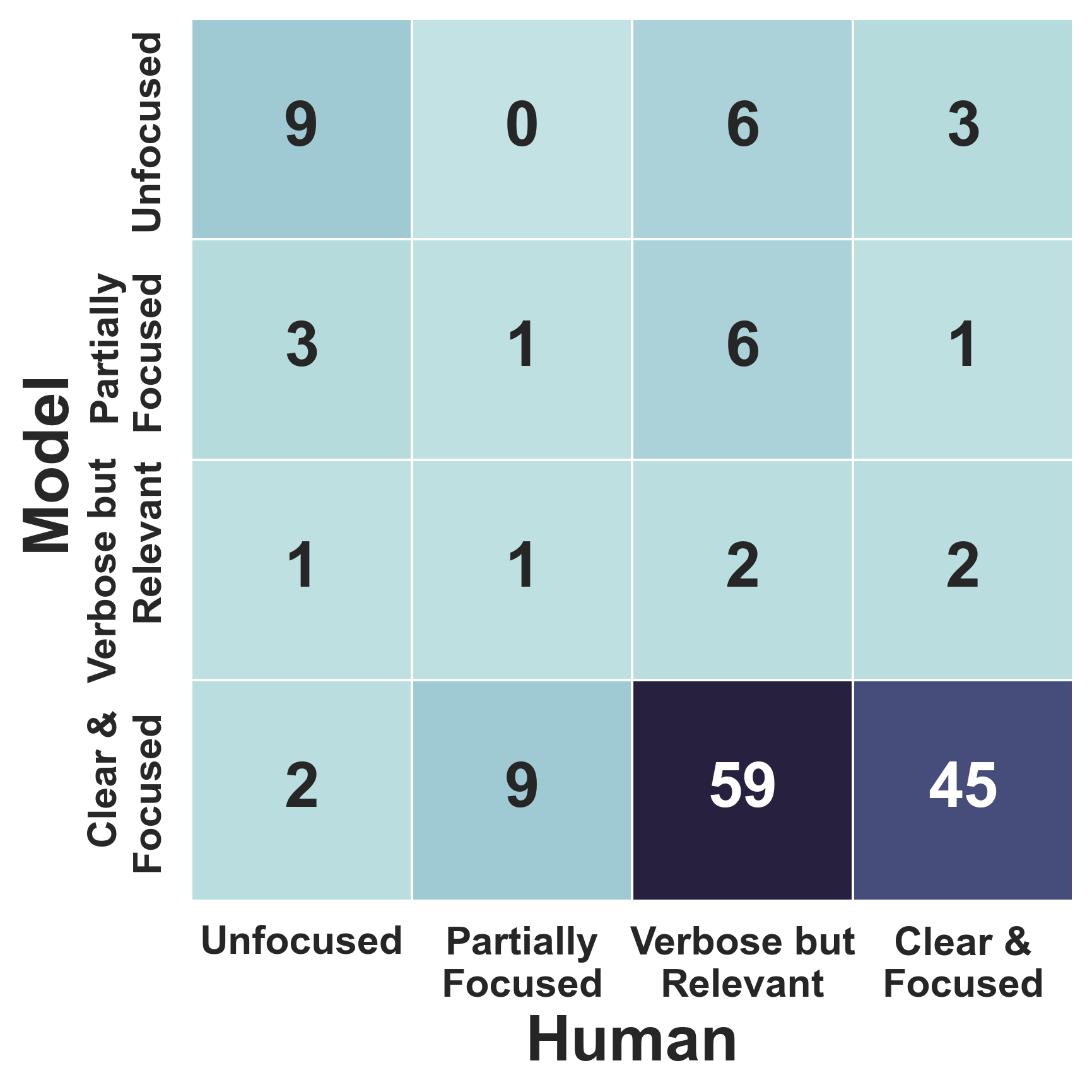}
     \caption{Relevance}\label{fig:rel}
 \end{subfigure}
 \begin{subfigure}[t]{.32\textwidth}
     \centering
     \includegraphics[width=\textwidth]{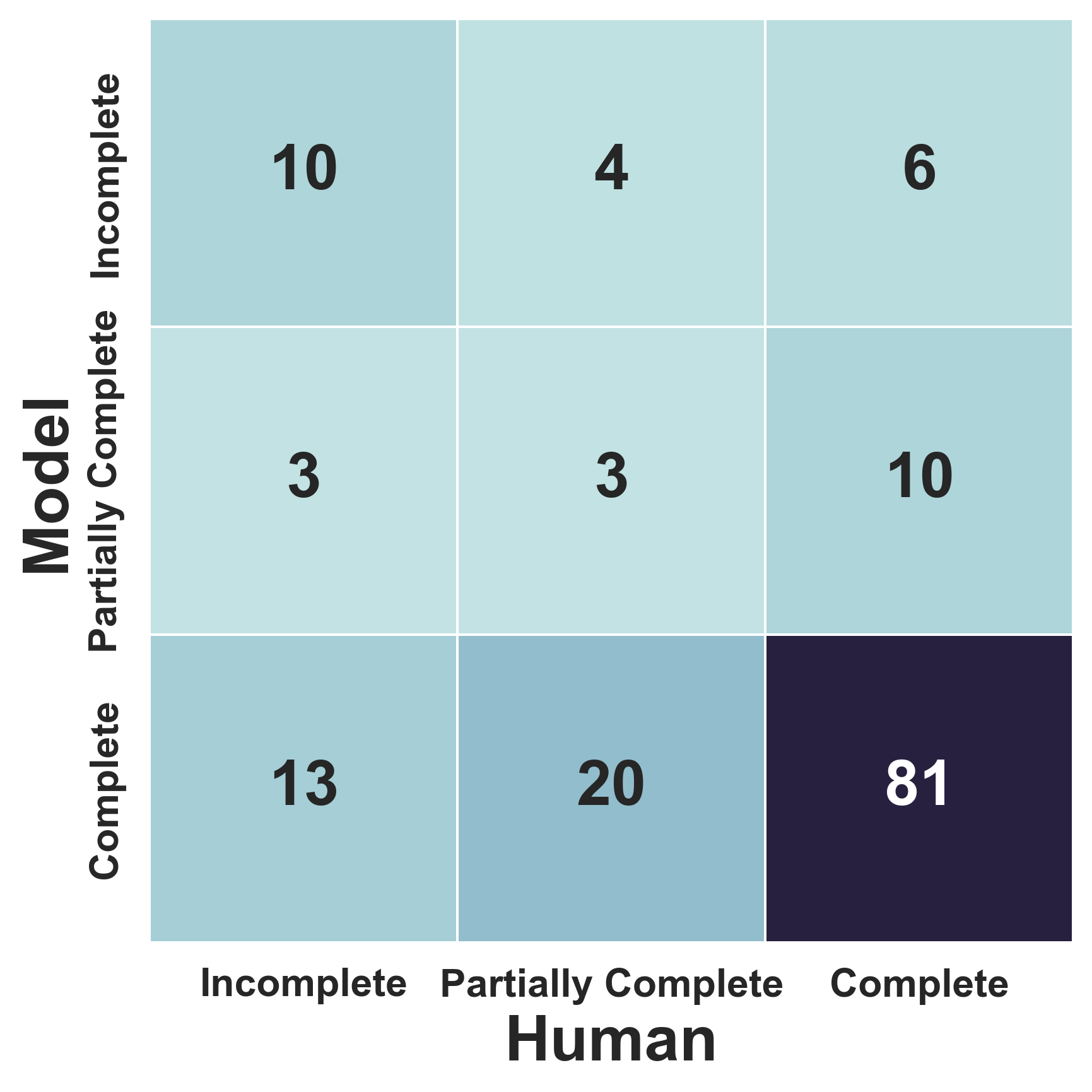}
     \caption{Completeness}\label{fig:comp}
 \end{subfigure}
 \caption{Model vs. Human agreement on (a) Correctness, (b) Relevance, and (c) Completeness. The model shown is GPT-o4-mini, and human labels are derived via majority vote.}
 \Description{}
 \label{fig:agreement}
\end{figure*}

\begin{figure*}[ht!]
 \centering
 \begin{subfigure}[t]{.32\textwidth}
     \centering
     \includegraphics[width=\textwidth]{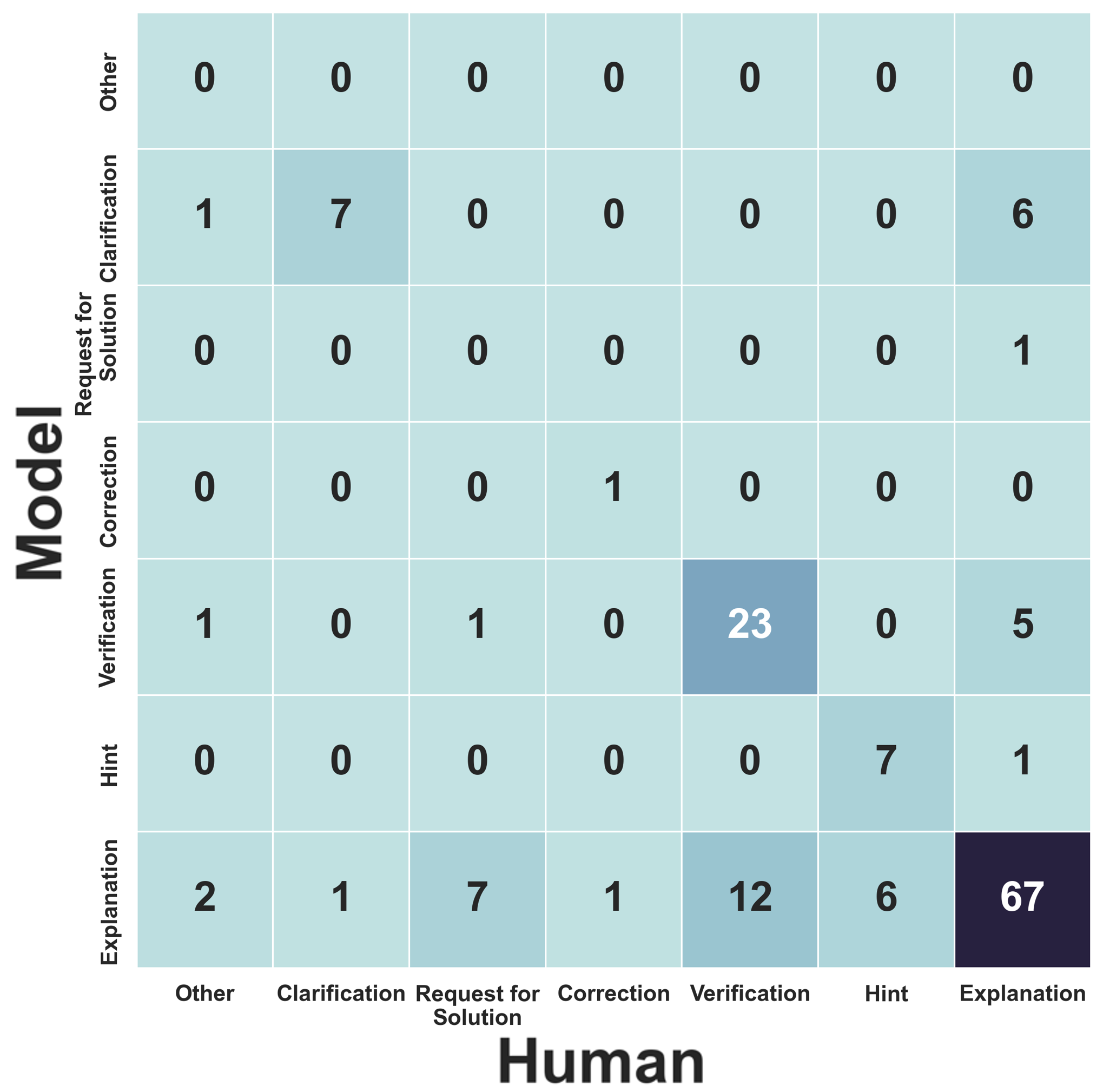}
     \caption{Expectations}\label{fig:exp}
 \end{subfigure}
 \begin{subfigure}[t]{.32\textwidth}
     \centering
     \includegraphics[width=\textwidth]{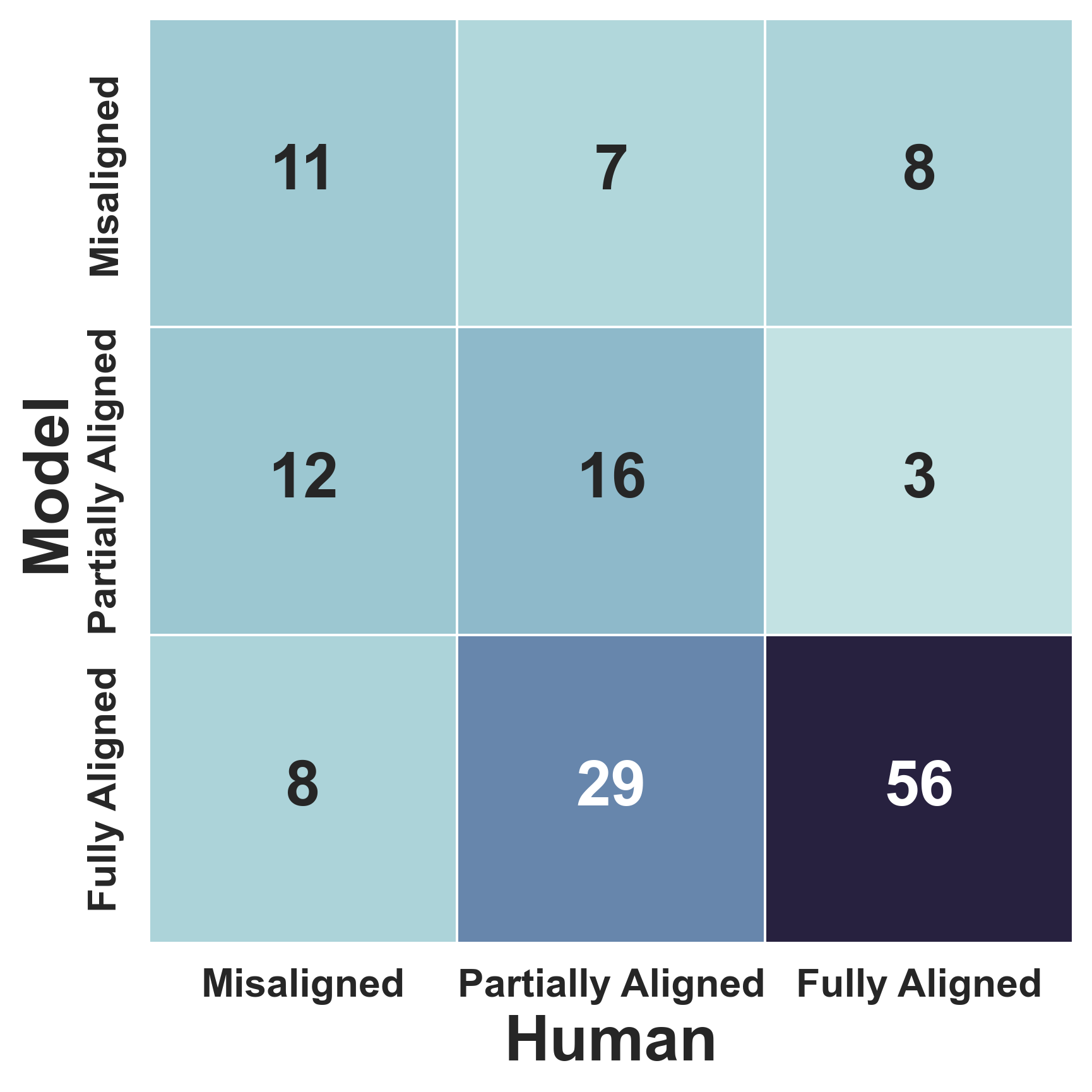}
     \caption{Expectation Alignment}\label{fig:exp_al}
 \end{subfigure}
 \begin{subfigure}[t]{.32\textwidth}
     \centering
     \includegraphics[width=\textwidth]{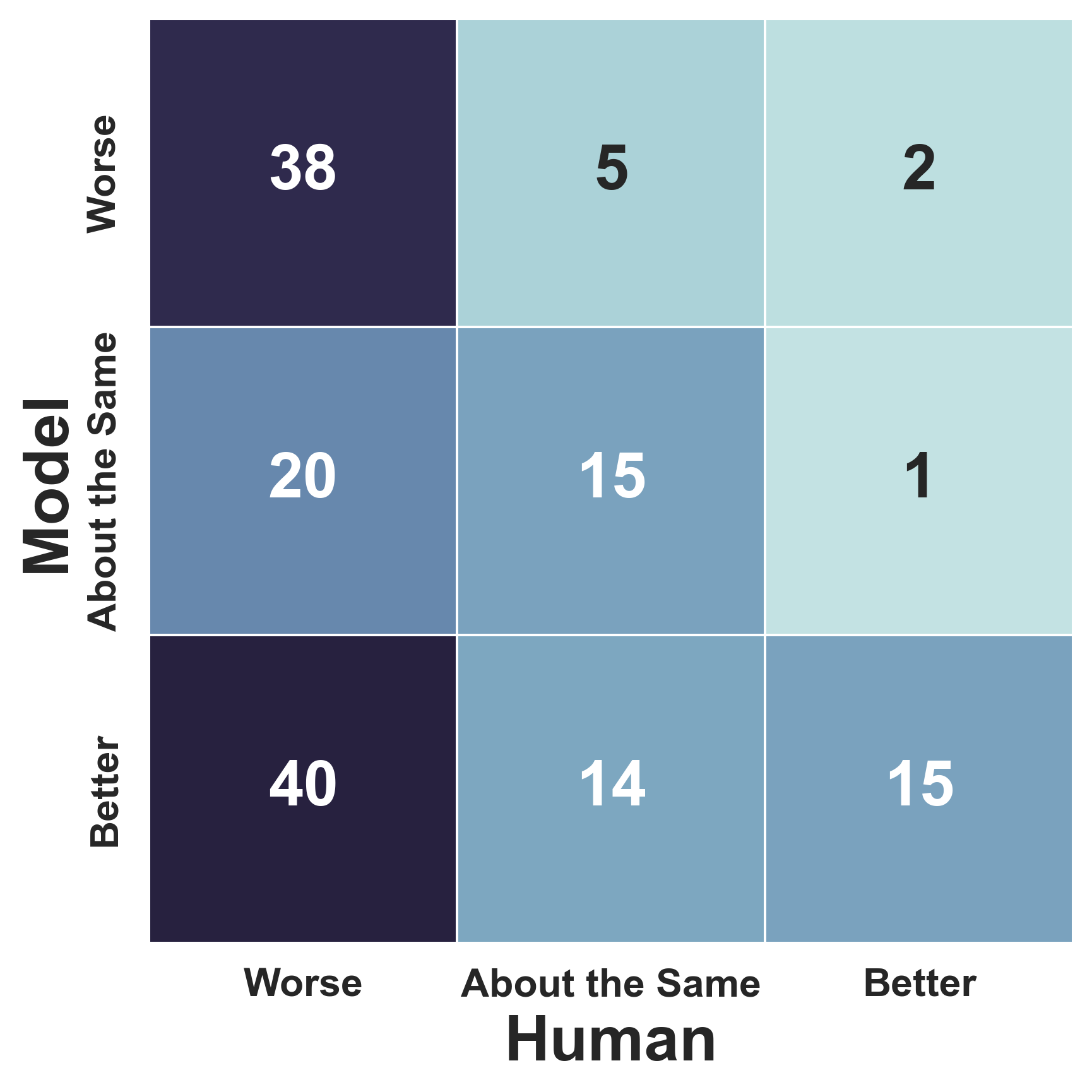}
     \caption{Comparison with Instructor}\label{fig:compar}
 \end{subfigure}
 \caption{Model vs. Human agreement on (a) Expectations, (b) Expectation Alignment, and (c) Comparison. The model shown is GPT-o4-mini, and human labels are derived via majority vote.}
 \Description{}
 \label{fig:agreement_2}
\end{figure*}

\subsection{RQ4: Dimensions of Responses Suitable for Automated Evaluation}

We compared automated evaluation metrics with expert instructor judgments to examine which response dimensions can be reliably assessed without human input. On average, the model’s assessments agreed with instructors in 55.3\% of cases across correctness, relevance, and completeness. Large discrepancies, such as assigning the highest possible score when instructors gave the lowest, were rare for objective dimensions like correctness. For example, as shown in Figure~\ref{fig:agreement}), only 6.7\% of cases were marked incorrect by the model but judged fully correct by instructors, and 4\% in the reverse. This indicates that automated metrics are reasonably dependable for correctness evaluation.

However, for more subjective dimensions such as relevance and completeness, the alignment was weaker. Moderate disagreements (e.g., one-point differences on a 3-point scale) were common, suggesting that automated methods struggled to capture the nuance between a ``good'' answer and a ``nearly perfect'' one. These subtler judgments often depended on pedagogical interpretation, whether an answer was sufficiently concise, whether an elaboration supported or distracted from learning goals, or whether an omission was critical.

\begin{figure*}[ht!]
 \centering

 \begin{subfigure}[t]{\textwidth}
     \centering
     \includegraphics[width=1.0\textwidth]{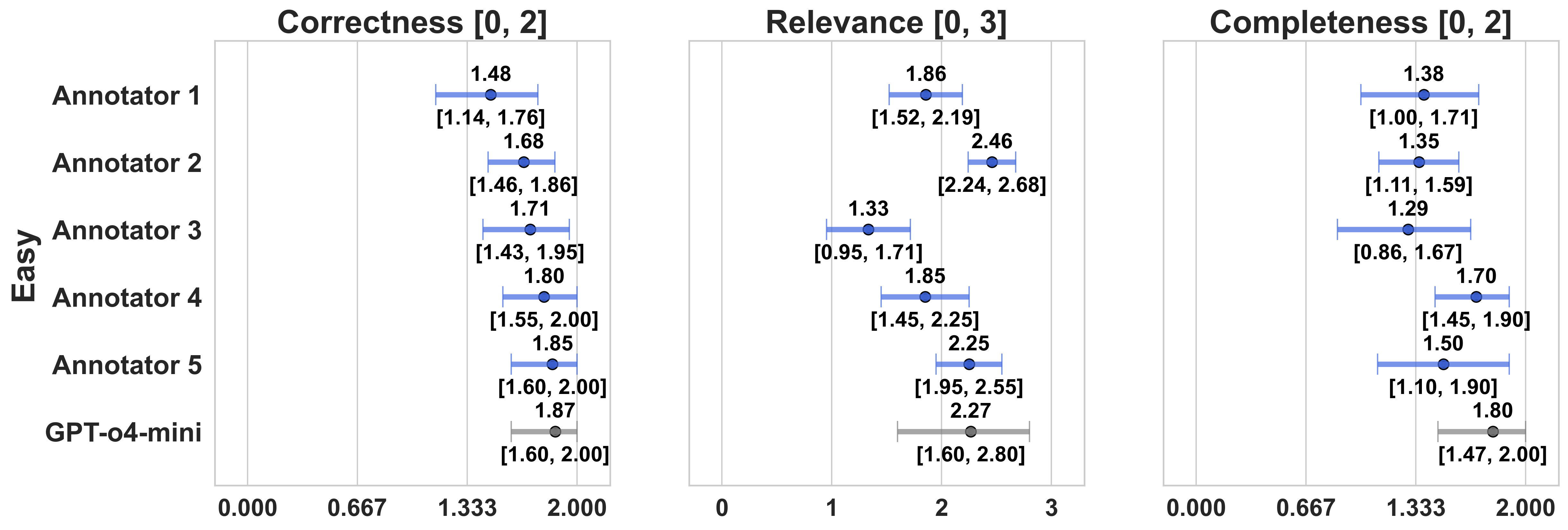}
     \caption{}\label{fig:easy2}
     \Description{}
 \end{subfigure}

 \begin{subfigure}[t]{\textwidth}
     \centering
     \includegraphics[width=1.0\textwidth]{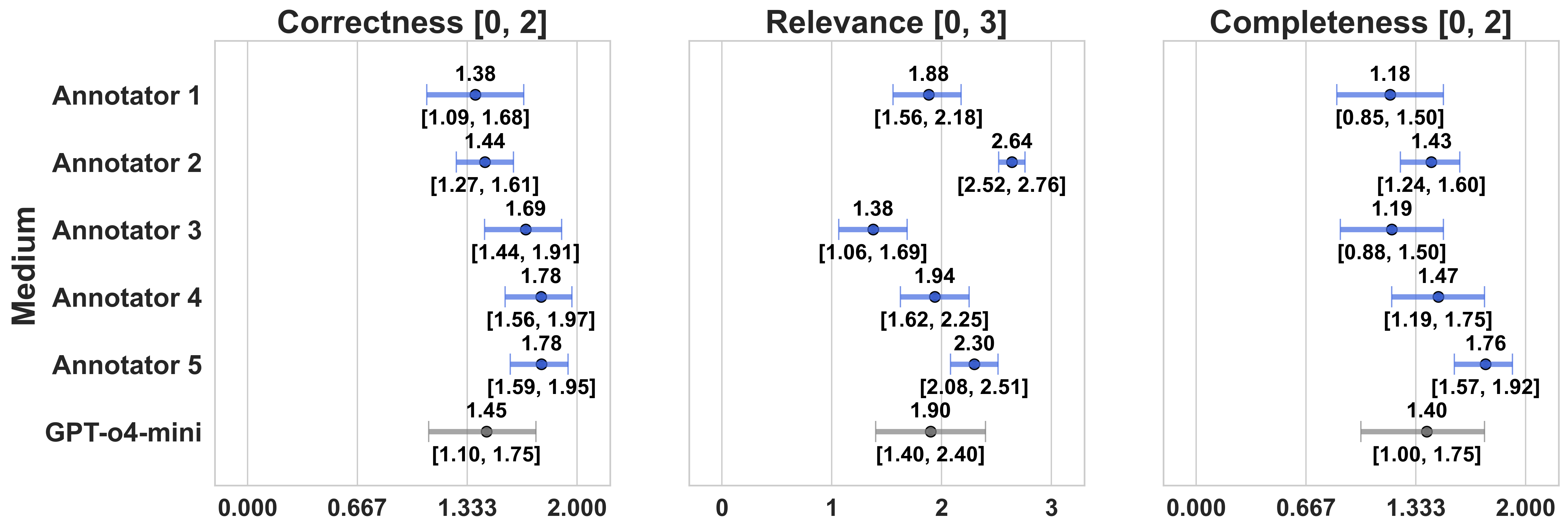}
     \caption{}\label{fig:medium}
     \Description{}
 \end{subfigure}

 \begin{subfigure}[t]{\textwidth}
     \centering
     \includegraphics[width=1.0\textwidth]{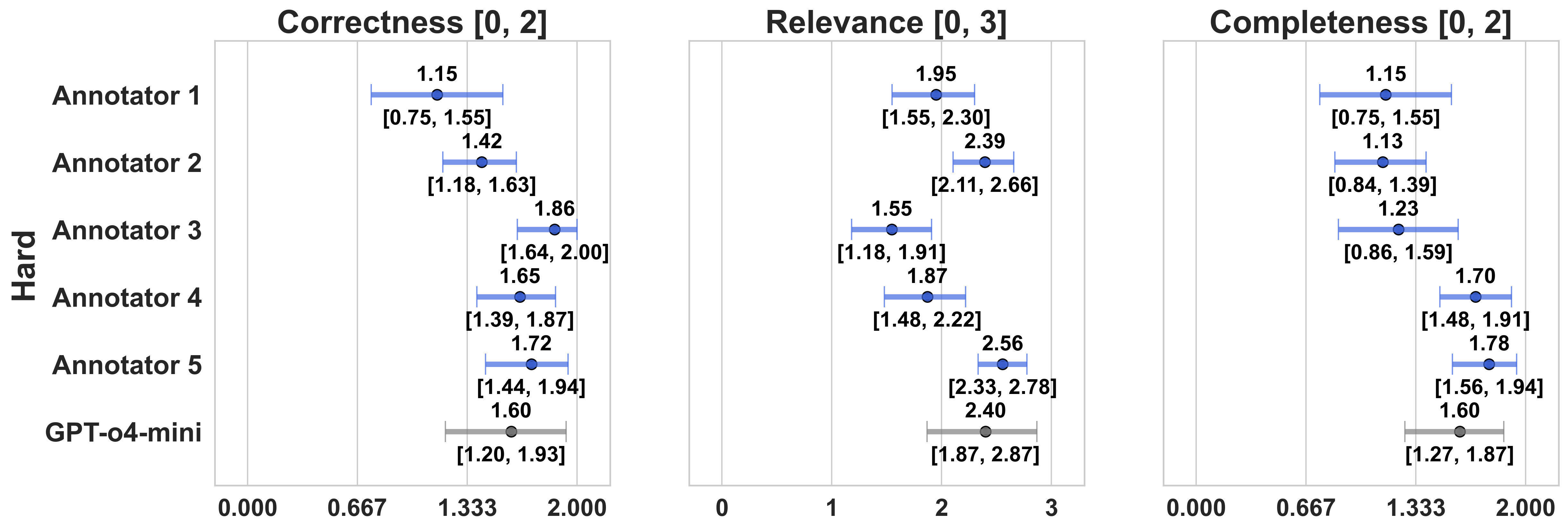}
     \caption{}\label{fig:hard2}
     \Description{}
 \end{subfigure}

 \caption{Comparison across annotators and automated evaluation with GPT-4o-mini. 95\% confidence intervals are computed via non-parametric bootstrap with 10000 resamples.}
 \Description{}
 \label{fig:mean_per_dimension_with_ci}
\end{figure*}

Overall, we observe that the model judgment would agree with at least one annotator, on average annotators and the model agree the most on easy and hard questions. On medium questions, the model is stricter, assigning lower scores than most annotators, particularly for Relevance and Completeness. Human expertise remains essential for evaluating pedagogical quality, intent alignment, and appropriateness of detail. The details are shown in Figure \ref{fig:mean_per_dimension_with_ci}.

Taken together, these findings suggest that automated evaluation is most effective for objective correctness judgments but less reliable for qualitative, pedagogical aspects of responses. As illustrated in Figure~\ref{fig:agreement_2}, the model diverged most strongly when answers were compared directly against an instructor’s version, reflecting the difficulty of encoding instructor-specific preferences.

This might imply that automated evaluation can reduce workload by filtering out clearly unacceptable responses, but human oversight remains essential for assessing qualities like relevance, tone, and pedagogical fit. Future designs may benefit from hybrid evaluation pipelines in which automated methods handle routine correctness checks, while instructors focus on higher-level dimensions that require contextual or pedagogical judgment. This supports existing research cautioning against over-reliance on model-based evaluation~\citep{abdurahman2024perils, chen-etal-2024-humans}.

\subsection{RQ5: Post-deployment Evaluation}

Starting at the beginning of the Fall 2025 semester, we deployed our model and the accompanying response pipeline on the course discussion forum, where it generated responses to students questions. We analyzed instructor actions, such as endorsements, edits, comments, and deletions, to assess real-world performance and determine how the system fit into the instructional workflow.

\begin{figure*}[ht!]
 \centering
 \begin{subfigure}[t]{.35\textwidth}
     \centering
     \includegraphics[width=\textwidth]{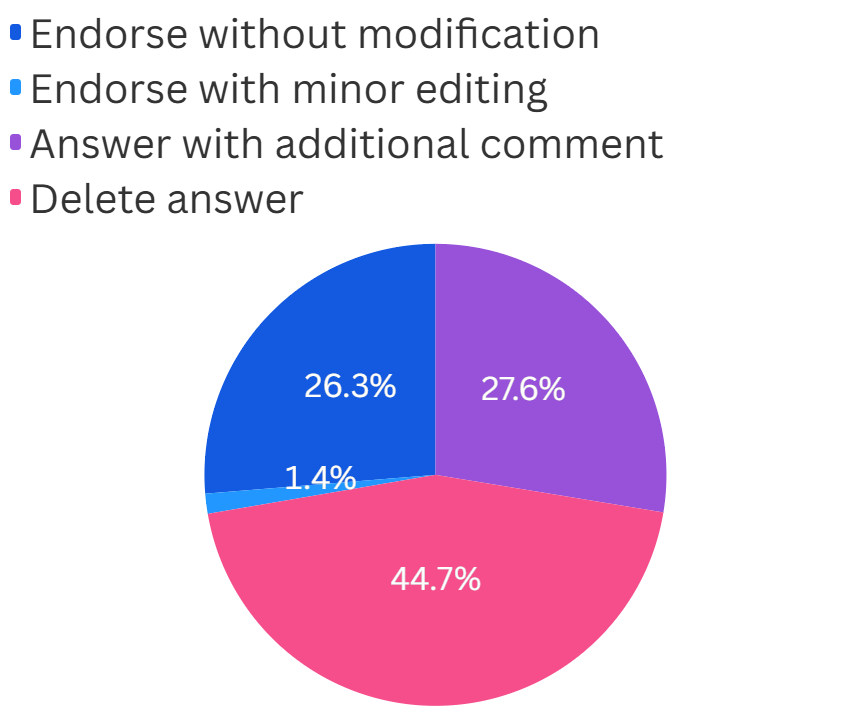}
     \caption{Instructor Actions}\label{fig:instructor_actions}
 \end{subfigure}
 \hfill 
 \begin{subfigure}[t]{.60\textwidth}
     \centering
     \includegraphics[width=\textwidth]{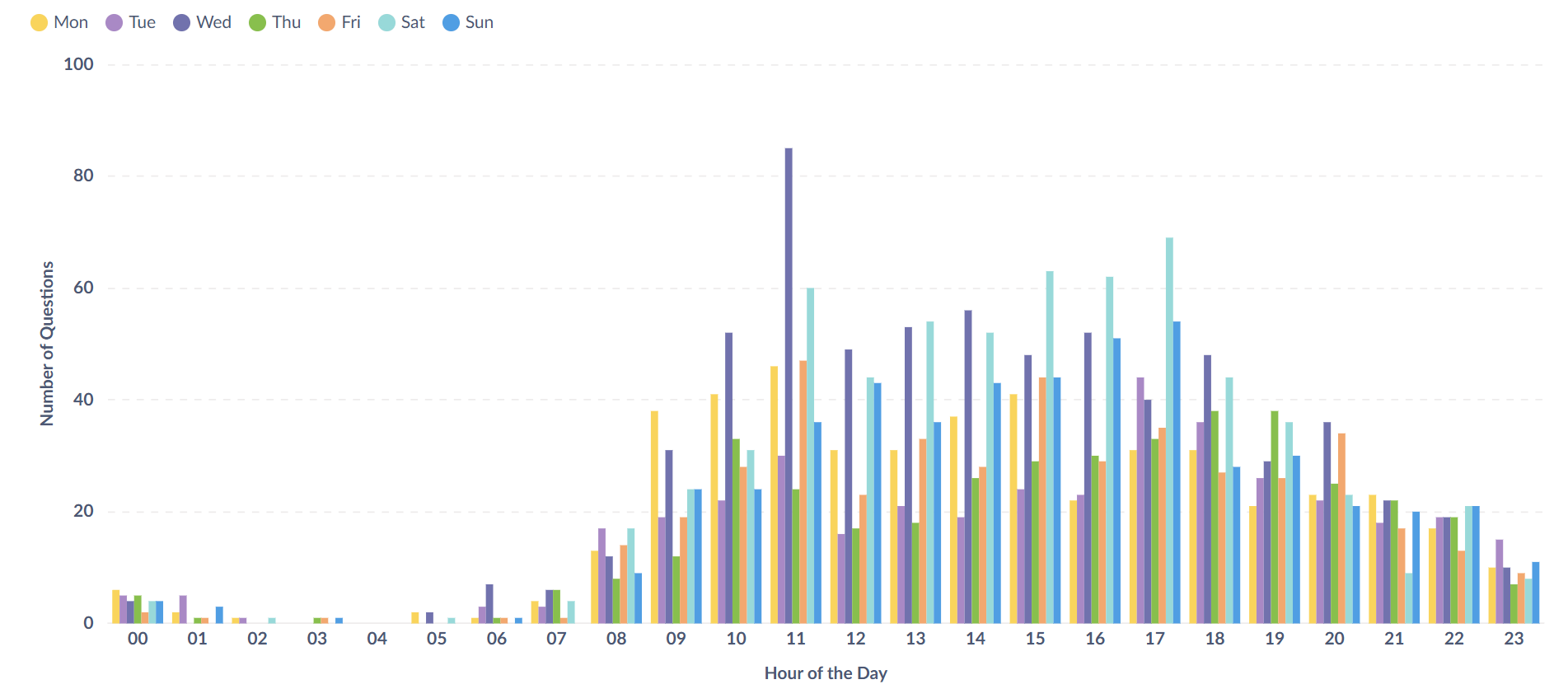}
     \caption{Hourly Volume}\label{fig:hourly_volume}
 \end{subfigure}
 \caption{Overview of model statistics: (a) Distribution of instructor actions taken on model-generated answers, and (b) Distribution of question volume by hour of day across weekdays.}
 \Description{}
 \label{fig:combined_stats}
\end{figure*}

\noindent \textbf{Usage Patterns:} As of November 25, the forum had 154 active student users, who posted an average of 3.7 questions each (range: 1–43). In total, students asked 569 questions since the beginning of the semester. Question volume was highly non-uniform across the semester, with pronounced spikes near homework and quiz deadlines. Posting activity was highest on Mondays, Wednesdays, and weekends. The weekends are the time when instructors are typically less available, underscoring the value of immediate automated responses. See Figure \ref{fig:hourly_volume}for the usage statistics. On average, 8.65 questions were asked per day, with daily counts ranging from 0 to 27.

\noindent \textbf{Instructor Actions:} Across all responses, the instructor endorsed 26.3\% without modification, indicating that the responses met instructional quality expectations. Only 1.4\% of responses were edited, the least frequent action. In interviews, the instructor explained that editing a model answer often required more effort than supplementing or rewriting it outright. Instead, the instructor preferred to augment the model’s response by adding a comment (27.6\% of cases), typically to clarify assumptions or provide alternative reasoning. In this case, the instructor reported that the model provided a useful baseline, allowing them to focus on deeper conceptual insight rather than drafting full replies. Student ``liked'' 22.4\% of kept responses. Finally, the instructor deleted 44.7\% of responses; these cases reflected replies judged as inappropriate or potentially misleading. Figure~\ref{fig:instructor_actions} summarizes these distributions.

\noindent \textbf{Instructor Interpretation:} Interview feedback revealed that deletions were not typically prompted by factual errors. More commonly, they reflected responses that were technically correct but pedagogically misaligned, such as overly complex explanations or answers addressing a different interpretation of the question than the student intended. We additionally examined temporal patterns in the proportion of model answers that were retained; this proportion varied across days, suggesting that model performance may differ by topic or question type, a hypothesis that needs further investigation. See Figure \ref{fig:combined_proportion} for more details.

Taken together, these findings show that the model can reliably handle routine questions at scale and emphasize the necessity of the instructor's oversight.

\subsection{RQ6: Students Perceptions of Model Responses}

After the system deployment, students were instructed that the model responses would appear first, almost immediately after they post a question, and would be marked as the model response. Later, typically within 24 hours, the instructor provided feedback on the model answers.

\noindent \textbf{Survey Participation and Usage:} Of 610 students enrolled in the class, 186 completed the survey. The forum has 154 distinct users. We restricted the analysis to the 105 respondents who had asked at least one content-related question on the course platform, which is 68.2\% of the forum users. Most reported using the question forum occasionally (77\%), with 21\% asking questions weekly and only 2\% asking daily. The most frequently selected categories were questions concerning the exercise statement (44.76\%), starting hints (36.19\%), continuation hints (36.19\%), and course-related questions (31.43\%), followed by result verification (26.67\%) and complete-solution requests (4.76\%). Percentages exceed 100\% cumulatively because multiple answers were permitted.

\begin{figure*}[t]
  \centering
  \includegraphics[width=0.9\textwidth]{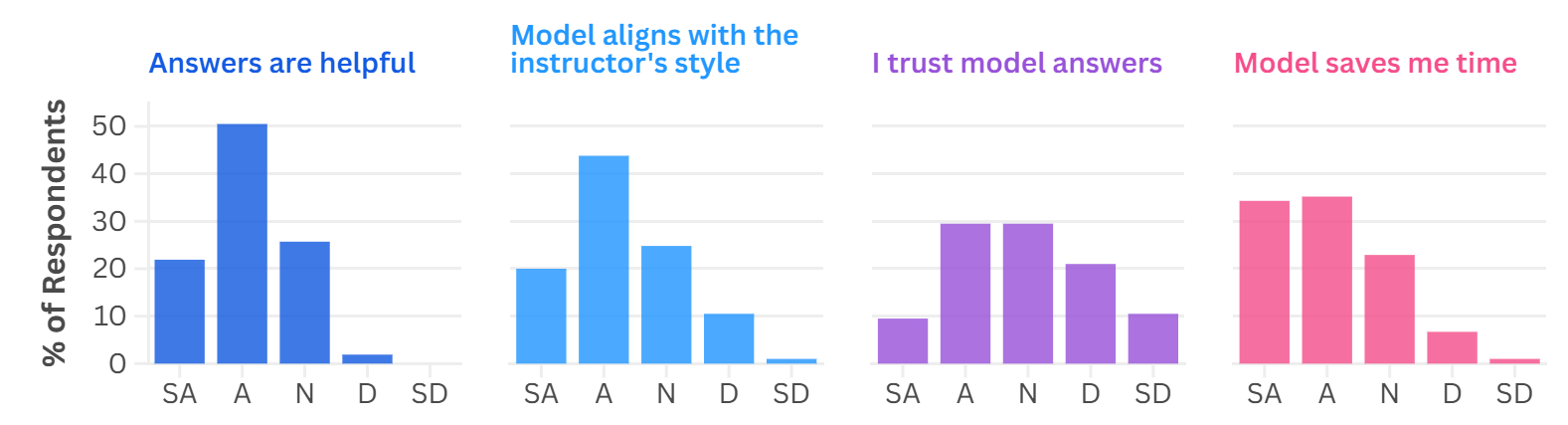}
  \caption{Student perceptions of the model’s responses, based on survey ratings (n = 105). (SA=Strongly Agree, A=Agree, N=Neutral, D=Disagree, SD=Strongly Disagree)}
  \Description{}
  \label{fig:student_form}
\end{figure*}

\noindent \textbf{Perceived Quality and Alignment:} Students generally viewed the model as aligned with their needs: 61\% reported that responses aligned well with their question intent, 20\% reported perfect alignment. Only 18.1\% rated alignment as ``fair.'' The majority found the answers helpful: 50.5\% agreed that model responses were helpful. 21.9\% strongly agreed, while 25.7\% were neutral. Students also noted stylistic consistency with the course instructor: 43.8\% agreed, 20\% strongly agreed, 24.8\% were neutral. The results are shown in Figure \ref{fig:student_form}.

\noindent\textbf{Trust and Reliance:} Trust levels were more divided. 39\% trusted the model enough to use its answers without always checking them (29.5\% agree, 9.5\% strongly agree). 31.5\% disagreed or strongly disagreed. This suggests that while students valued the model, some need answers verification which is consistent with instructor interviews and open feedback from students.

\noindent \textbf{Perceived Benefits:} Students widely appreciated the immediacy of responses: 69.5\% (35.2\% agree, 34.3\% strongly agree) reported that immediate answers saved them time. 80\% expressed interest in having similar system in other courses.

\noindent \textbf{Qualitative Feedback.} Among 54 respondents who provided written comments, several themes emerged:

\begin{itemize}
    \item Alignment with course materials ($\approx$ 30\%): Students valued that the model used familiar notation, examples, and terminology, which reduced confusion compared to general-purpose tools.
    \item Immediate but verified support ($\approx$ 29.6\%): Many appreciated receiving a quick answer followed by instructor confirmation.
    \item Preference for detailed step-by-step explanations ($\approx$ 35.2\%): Some students found the model’s elaboration more helpful than typical short instructor replies.
    \item Preference for succinct hints ($\approx$ 32.7\%): Others preferred brief guidance. This mirrors instructor concerns about pedagogical appropriateness and highlights potential benefits of personalization.
    \item Preference for conversational interfaces: Several students asked for a chat-style interface separate from the question forum, suggesting a desire for more exploratory, iterative, and personalized help.
\end{itemize}

Here are a few examples of student comments:

\begin{quote}
\textit{``The model prevents me from getting stuck on an exercise and helps me understand conceptual errors in my reasoning. It also ensures that it is a reliable source and saves time. There is no need to wait for a professor or classmate to respond.''} (P12)
\end{quote}

\begin{quote}
\textit{``Since it is entirely based on the course website, it never produces answers out of nowhere, which is good.''} (P18)
\end{quote}

\begin{quote}
\textit{``It helps me understand new concepts and better understand an exercise without having to wait for a teacher's response, which sometimes provides fewer details than the language model.''} (P27)
\end{quote}

\begin{quote}
\textit{``Because the answers are based only on our course materials, they match what we have learned. With ChatGPT, I always have to provide the course context so it knows my level.''} (P41)
\end{quote}

Overall, students reported that the model provided timely, course-aligned assistance that complemented instructor support. However, preferences varied substantially, especially regarding explanatory depth, pointing toward opportunities for adaptive or personalized response strategies.

\section{Design Implications for AI in Large Courses}

Our quantitative, qualitative, and behavioral analyses yield several design implications for building effective, safe, and context-aware AI support systems for large-enrollment courses:

\noindent \textbf{Lightweight course-specific models are sufficient for pedagogical utility.} A lightweight, course-tailored model can reach a level of accuracy that is pedagogically useful. 

\noindent \textbf{Course alignment matters for students.} Students strongly valued that, for resposnes, the model uses familiar notation, examples, and contextualize with the lecture notes. 

\noindent \textbf{Instructor feedback and iterative adaptation are essential.} Active involvement of instructors during model development helps to detect model's subtle instructional issues. Effective deployments should remain flexible, allowing updates to the model and pipeline in response to instructor and student needs, as well as improvements in model capabilities.

\noindent \textbf{Hybrid human–AI system improves trust and mitigates risks.} Instructor verification not only corrects edge cases and pedagogical mismatches but also helps students develop awareness of model limitations.

\noindent \textbf{There is a demand for response detail personalization.} Students differ in their preferences for concise hints versus full explanations. Effective systems should calibrate response depth based on both question intent and individual learner needs.

\noindent \textbf{High student trust requires transparency about limitations.} Although students trust the model more because it is trained on course materials, it still produces subtle or pedagogically inappropriate responses. Clear communication of known limitations is essential.

\noindent \textbf{Pedagogical appropriateness is the key remaining challenge.} Most errors were not factual but instructional: misinterpreting question intent, giving overly complex or trivial reasoning, or providing mismatched levels of detail. Future work must prioritize modeling pedagogical intent and instructional strategy, not only correctness.





\section{Conclusion}

In this case study, we examined the development, evaluation, and deployment of a course-tailored generative AI model for supporting a large-enrollment Calculus I course. Through close collaboration with the instructor, we created a course-specific dataset, a representative evaluation benchmark, and human-centered assessment criteria that guided evaluation. A lightweight, locally deployable model fine-tuned on 2,738 real student–instructor interactions achieved strong performance, especially on the medium-difficulty (most common questions in the course).

Our findings show that while the model produced largely accurate and contextually grounded responses, it could not reliably replicate the pedagogical nuance and intent-sensitive reasoning of human instructors. Errors were often subtle rather than overt, underscoring the importance of situating AI within a hybrid human–AI workflow rather than positioning it as an autonomous tutor. Instructor oversight not only mitigated risks but also increased student trust in the system, with many students reporting appreciation for both the immediacy of AI-generated answers and the subsequent human verification.

Students overall valued the model’s alignment with course terminology and examples, though preferences diverged regarding the desired level of detail: some favoring step-by-step explanations and others preferring concise hints. This highlights a future need for finer-grained personalization, not only at the course level but potentially at the learner level as well.

This work contributes empirical evidence, a road map, and design implications for integrating generative AI responsibly into STEM instruction. Beyond demonstrating the value of small, course-specific models, it emphasizes that effective educational AI must prioritize pedagogical appropriateness, transparency, and human oversight. These findings highlight a path toward AI systems that complement, rather than replace, the instructional expertise essential to student learning.

\section*{Limitations}

This work has several limitations that inform directions for future research.

First, our study centers on a single course context, Calculus I at one institution, where both the dataset and evaluation criteria were grounded in the practices of a specific instructor and a small group of experienced annotators. This narrow scope allowed us to conduct a deeply contextualized, human-centered case study, but it limits the generalizability of our findings to other disciplines where answers are less well-defined. While mathematics offers advantages for benchmarking due to objective correctness, future work should explore more interpretive fields where pedagogical alignment and explanation style may play an even larger role.

Second, our model was deliberately designed to be lightweight and deployable within institutional infrastructures, prioritizing data security and cost efficiency. This design choice enhances accessibility but constrains scalability and raw performance compared to larger cloud-based systems. This highlights a broader tension in educational AI design: how to balance local control, privacy, and inclusivity with the capabilities of state-of-the-art models.

Third, this work does not explore advanced approaches for automated evaluation. We rely on a direct application of GPT-4o for automated judgments and compare the results against human annotations. While this setup allows us to validate the feasibility of LLM-based evaluation, it does not fully address the scalability and cost limitations associated with human annotation. Future work could investigate more sophisticated automated evaluation strategies to reduce reliance on expensive and time-consuming manual labeling. Promising directions include fine-tuning dedicated classifiers on human-annotated data, calibrating LLM evaluators using small but high-quality annotation sets, and developing human-in-the-loop pipelines in which annotators focus only on examples where the model’s confidence is low or predictions are inconsistent. Such approaches could improve both the efficiency and reliability of automated evaluation at scale.

Finally, we collected students’ perceptions of the model-generated answers, including their satisfaction, trust in the responses, and attitudes toward having a model provide answers that are later endorsed or supplemented by instructors. Although the surveys were fully anonymous, responses may still be subject to social desirability bias. For example, students may overreport behaviors such as consistently verifying model outputs, even unintentionally.

\section*{Ethics Statement}
This study was conducted in accordance with institutional ethical guidelines and was approved by our IRB; details are withheld to maintain anonymity. All human annotators involved in the evaluation were university-level instructors with extensive experience teaching Calculus I or students of the course. They participated voluntarily and were informed of the study’s goals and procedures. No personally identifiable information was collected during annotation or evaluation.

Questions posted by students on the question forum are initially anonymized, and we don't collect any identifying information, such as the year of the question. All data originated from a single university's course portal and was used with the data owner's permission. Student questions were historical, not real-time, and no active participation from students was solicited for this study.

We emphasize that the AI model is designed to assist, not replace, instructors, providing scalable instructional support in large courses. Thus, there are no potential risks for students, since, as the final result, they will be getting either instructor-verified answers from the model or instructor-provided answers if the model answer is not satisfactory.

In terms of environmental impact, fine-tuning required 3 hours on 16 H100 GPUs for the course data and about 30 hours on the same hardware for the OpenMath220k dataset. Altogether, the experiments consumed approximately 1,900 GPU-hours, resulting in an estimated 73 kg of CO$_2$ emissions. We prioritized the use of lightweight, cost-efficient models to reduce computational overhead and support deployment on local infrastructure.
\begin{acks}

We thank the reviewers for their thoughtful comments and helpful suggestions, which significantly contributed to improving the framing and clarity of this paper. We are grateful to the course instructors who participated in the model annotation process.

We thank the students who actively contributed to the development of the platform: Hadrien Sevel (Master’s student, ETH Zurich), Loïc Misenta (Master’s student, EPFL), and Maxence Espagnet (IN student, EPFL).

We gratefully acknowledge the financial support provided by Simone Deparis (CePro, EPFL); the DRIL project “A language model for a teaching platform in Analysis 1” and Patrick Jermann (CEDE, EPFL), and François Genoud (CMS, EPFL). For technical infrastructure and support, we thank Khadidja Malleck and Matéo Müller (RCP, EPFL).

Finally, we gratefully acknowledge the support of the Swiss National Science Foundation (Grant No. 215390), Innosuisse (PFFS-21-29), the EPFL Center for Imaging, Sony Group Corporation, a Meta LLM Evaluation Research Grant, and the Swiss National Supercomputing Center (CSCS) through an infrastructure engineering and development project.

\end{acks}

\bibliographystyle{ACM-Reference-Format}
\bibliography{sample-base, custom}

\appendix
\section{Dataset Details}\label{app:datastat}
In this section, we provide information about the dataset statistics in terms of the token numbers, the approach for the question augmentation, and details on the model, which we used for translating our dataset from French to English.

\subsection{Dataset Statistics}

Table \ref{tab:token-stats} shows the token statistics for our dataset.

\begin{table}[htbp]
\small
\centering
\caption{Token statistics for context, questions, and non-augmented original answers in French and English datasets based on the GPT-4o tokenizer.}
\begin{tabular}{lccc}
\toprule
\textbf{Language} & \textbf{Context} & \textbf{Question} & \textbf{Answer} \\
& mean / max & mean / max & mean / max \\
\midrule
French & 891 / 5320 & 56 / 651 & 90 / 725 \\
English & 864 / 5214 & 52 / 634 & 82 / 692 \\
\bottomrule
\end{tabular}
\label{tab:token-stats}
\end{table}

The augmented answers have a total mean length of 1,337 tokens. The lecture notes only are in total 860k tokens. The whole dataset including context-chunks, questions and augmented answers is about 7.6M tokens in total.

The OpenR1-Math-220k dataset in the default split contains 93 733 unique questions and answers (with reasoning). In total, this resulted in 527M tokens with a mean length of 5626, lower quartile at 2764 and upper quartile of 7653 tokens.


\subsection{Dataset Augmentation}\label{app:augmentation}

For the augmentation of answers, we use the DeepSeek-R1-Distill-Qwen-32B model. 

Here is an example of a Q\&A pair without augmentation.

\textbf{Context:}

\begin{quote}
    Let $((a_n)_{n\geq 1})$ be the sequence defined by $(a_n=(-1)^n+\frac{1}{n})$, and
let $(A=\{a_1,a_2,a_3,\dots\})$.
\end{quote}

\textbf{Question:} 
\begin{quote}
Can we use the Bolzano-Weierstrass theorem to find the infimum and supremum of this sequence?
\end{quote}

\textbf{Instructor Answer:}
\begin{quote}
\noindent
The Bolzano--Weierstrass Theorem is not a result that allows you to find things; it is an existence theorem. It simply states, under the right hypotheses, that \textit{there exists a subsequence} with such properties, but it does not tell you exactly how to find it. If you look at the proof, you will see that this subsequence can be constructed using a superior or inferior limit, but nothing more concrete than that.
\end{quote}

Here is an example of the same answer after the augmentation: 

\begin{quote}
\noindent
To determine the infimum and supremum of the sequence \( a_n = (-1)^n + \frac{1}{n} \), we can analyze its behavior directly without relying on the Bolzano-Weierstrass theorem. Here is a step-by-step explanation:

\begin{itemize}
    \item Understanding the Sequence: The sequence alternates due to the \( (-1)^n \) term. For even \( n \), \( a_n \) is positive; for odd \( n \), it is negative. The term \( \frac{1}{n} \) decreases as \( n \) increases, approaching zero.
    
    \item Boundedness: The sequence is bounded because \( (-1)^n \) alternates between \(-1\) and \(1\), and \( \frac{1}{n} \) is always positive and less than or equal to \(1\). Thus, the sequence is bounded between \(-1\) and \(2\).
    
    \item Limit Points:
    \begin{itemize}
        \item For even \( n \), \( a_{2k} = 1 + \frac{1}{2k} \) approaches \(1\) as \( k \to \infty\).
        \item For odd \( n \), \( a_{2k+1} = -1 + \frac{1}{2k+1} \) approaches \(-1\) as \( k \to \infty\).
    \end{itemize}
    
    \item Supremum and Infimum:
    \begin{itemize}
        \item The supremum (least upper bound) is the maximum value of the sequence, which occurs at \( n = 2 \), giving \( a_2 = 1.5 \).
        \item The infimum (greatest lower bound) is the minimum value of the sequence, which occurs at \( n = 1 \), giving \( a_1 = -1 \).
    \end{itemize}
    
    \item Conclusion: While the Bolzano--Weierstrass theorem guarantees the existence of convergent subsequences for bounded sequences, it is not necessary here. The structure of the sequence allows us to directly determine the infimum and supremum by examining its terms.
\end{itemize}

\noindent Final Answer: The infimum of the sequence is \(-1\), and the supremum is \(1.5\).
\end{quote}

The prompt used for the augmentation is provided in Appendix \ref{app:prompts}.

\subsection{Model for Translation}

For translation from French to English, we use the GPT-4o (version 2024-08-06). A subset of translations was manually verified by a speaker fluent in both French and English.

The prompt used for the translation is provided in Appendix \ref{app:prompts}.

\section{Model Implementation}\label{app:model}
In this section, we summarize details on base models and the finetuned models. Along with their performance on benchmarks.

\subsection{All Models from Experiments}
We selected a diverse set of models to balance mathematical reasoning, context length, and language coverage:
Llama-3.1-8B-Inst (Huggingface ID: meta-llama/Llama-3.1-8B) is very versatile and well-aligned with system prompts, with the added benefit of multilingual support, but it falls short on specialized math knowledge. 

Mathstral-7B (Huggingface ID: mistralai/Mathstral-7B-v0.1) brings strong French and English math capabilities (could be useful since our course materials are in French), but it has weaker math performance compared to similar models of the same size. 

Qwen-2.5-Math-7B (Huggingface ID: Qwen/Qwen2.5-Math-7B-Instruct), is tuned specifically for math benchmarks, delivers very rigid and high-accuracy solutions in English and Chinese, but lacks French support and only offers a 4k token context.

OpenR1-Qwen-7B (Huggingface ID: open-r1/OpenR1-Qwen-7B) builds on Qwen’s strengths with superior math reasoning and a 32k extended window, but remains constrained to English and Chinese. 

Gemma-3-12B-Inst (Huggingface ID: google/gemma-3-12b-it), is not the strongest math model, but it is a promising multimodal, multilingual candidate. Its French support and potential for image understanding could be useful for questions with screenshots or diagrams. 

DeepSeek-R1-Distill-14B (Huggingface ID: deepseek-ai/DeepSeek-R1-Distill-Qwen-14B) stands out as the top math reasoning model at this scale; the only potential limitation is the language capabilities.

These eight models span four families: Llama, Mistral, Qwen, and Gemma. For a complete breakdown of training epochs and context-window sizes, see Table \ref{tab:all-models}. Each model is publicly available on the Hugging Face Hub under the identifiers listed in the table. The list of all models and datasets is available on \href{https://huggingface.co/collections/Jeremmmyyyyy/ai-meets-mathematics-education}{Huggingface}.

\begin{table*}[t!]
\centering
\small
\caption{Overview of model configurations. \textsuperscript{*} in the context window column for Open-R1-QA indicates that the base Qwen-2.5-Math-7B model was extended from a 4k to a 32k token context window. The Gemma-QA\_OpenMath220k model was fine-tuned for 3 epochs on the OpenMath220k dataset, followed by 3 epochs on the course dataset. All other models were only fine-tuned on course data.}
\begin{tabular}{lccc}
\toprule
\textbf{Model} & \textbf{Model Family} & \textbf{Epochs} & \textbf{Context Window}\\
\midrule
Llama-3.1-8B-Inst & llama & - & 128k \\
Mathstral-7B& mistral & - & 32k \\
Qwen-2.5-Math-7B & qwen & - & 4k  \\
OpenR1-Qwen-7B & qwen & - & 32k \\ 
Gemma-3-12B-Inst & gemma & - & 128k \\
DeepSeek-R1-Distill-14B & qwen & - & 128k\\
\midrule
Qwen-2.5-QA & qwen & 10 & 4k  \\ 
Gemma-QA\_OpenMath220k & gemma & 3+3 & 128k  \\
Open-R1-QA & qwen & 3 & 32k\textsuperscript{*}\\
DeepSeek-R1-QA   & qwen & 3 & 128k  \\
\bottomrule
\end{tabular}
\label{tab:all-models}
\end{table*}

All trained models were evaluated on established, challenging math and STEM benchmarks: AIME24, AIME25, MATH\_500,\\ GPQA:Diamond, and UMATH. For the first four, we used the standard implementation guidelines and integrated them into custom Lighteval tasks \cite{lighteval}. For UMATH, we selected the 900 text problems (out of the full 1,200), and rather than employing a larger LLM to grade responses, we applied the same LaTeX-based extraction metric as for the MATH\_500 benchmark directly to the model outputs. Models' performance on benchmarks is presented in Table \ref{tab:math-results}.

Our best performing model, DeepSeek-R1-QA, was finetuned for three epochs on the augmented question–answer pairs (see Appendix \ref{app:augmentation} for details). 
We used a context window of 16k tokens, a global batch size of 1 with four gradient accumulation steps, and a constant learning rate schedule with a 20\% linear warmup. Optimization was performed with paged 8-bit Adam and a maximum gradient norm of 1.0. We trained the model in its full 16-bit (bfloat16) precision without sequence packing. Under these settings, training required approximately three hours on 16 H100 GPUs. 

\begin{table*}[t!]
\centering
\small
\caption{Performance of the base models and the fine-tuned model on the selected math benchmarks. ``QA'' means that the model was finetuned on our course data, ``OpenMath220k'' means that it was also finetuned on the OpenR1-Math-220k dataset.  A dash (–) indicates that the model was not evaluated on that metric. We highlight the best results across the base models and our models with \textbf{bold}.}
\begin{tabular}{lccccc}
\toprule
\textbf{Model} & \textbf{AIME24} & \textbf{AIME25} & \textbf{MATH\_500} & \textbf{GPQA:Diamond} & \textbf{U\_math} \\
\midrule
Llama-3.1-8B-Inst & 0.066 & 0.000 & 0.460 & 0.369 & 0.144 \\
Mathstral-7B            & 0.033 & 0.066 & 0.486 & 0.298 & 0.146 \\
Qwen-2.5-Math-7B & 0.133 & 0.133 & 0.834 & 0.272 & 0.302 \\
OpenR1-Qwen-7B & 0.533 & 0.400 & 0.932 & 0.424 & 0.504 \\ 
Gemma-3-12B-Inst               & 0.267 & 0.200 & 0.860 & 0.349 &   0.342   \\
DeepSeek-R1-Distill-14B        & \textbf{0.667} & 0.533 & 0.958 & 0.611 &  0.541   \\
GPT 4o                               & 0.167 & 0.060 & 0.738 & 0.545 &   –   \\
DeepSeek-R1-671B& \textbf{0.667} & \textbf{0.633} & \textbf{0.970} & \textbf{0.727} &   \textbf{0.578}   \\
\midrule
Qwen-2.5-QA & 0.066 & 0.033 & 0.562 & 0.308 & 0.170 \\ 
Gemma-QA\_OpenMath220k & 0.133 & 0.200 & 0.626 & 0.323 &   0.202   \\
Open-R1-QA & 0.367 & 0.233 & 0.862 & 0.298 & 0.359 \\ 
DeepSeek-R1-QA   & \textbf{0.633} & \textbf{0.400} & \textbf{0.934} & \textbf{0.601} &   \textbf{0.461}   \\
\bottomrule
\end{tabular}
\label{tab:math-results}
\end{table*}

\subsection{Model Adjustments after Deployment}\label{app:adjustements}

We incorporated the following add-ons to our model pipline.

For image and text recognition, we employed a vision-enabled Mistral model (HuggingFace ID: mistralai/Mistral-Small-3.2-24B-Instruct-2506). Although this model is not optimized specifically for OCR or text-to-image tasks, it was selected due to its availability through our university’s shared infrastructure. The model was used to generate detailed textual descriptions of the uploaded images and any embedded text, enabling these descriptions to be appended to the original question before being forwarded to the specialized mathematics model.

For question classification and answer refinement, we utilized a GPT-OSS model (HuggingFace ID: openai/gpt-oss-120b). This model provides strong general-purpose text understanding and supports tool usage, which is essential for reliable structured classification via a JSON-based parser that allows retries and guarantees a single category selection. Its summarization and text-processing abilities also made it suitable for refining and clarifying outputs from the specialized mathematics model. While GPT-OSS is a large model that typically requires professional-grade hardware, we also evaluated a more lightweight alternative, Qwen-3 (HuggingFace ID: Qwen/Qwen3-30B-A3B). This model can be used interchangeably in the pipeline and operates efficiently on consumer-grade hardware, including GPUs with 16 GB of memory, while still supporting tool-calling functionalities.

\subsection{Weekly model usage}

\begin{figure*}[t]
  \centering
  \includegraphics[width=0.7\textwidth]{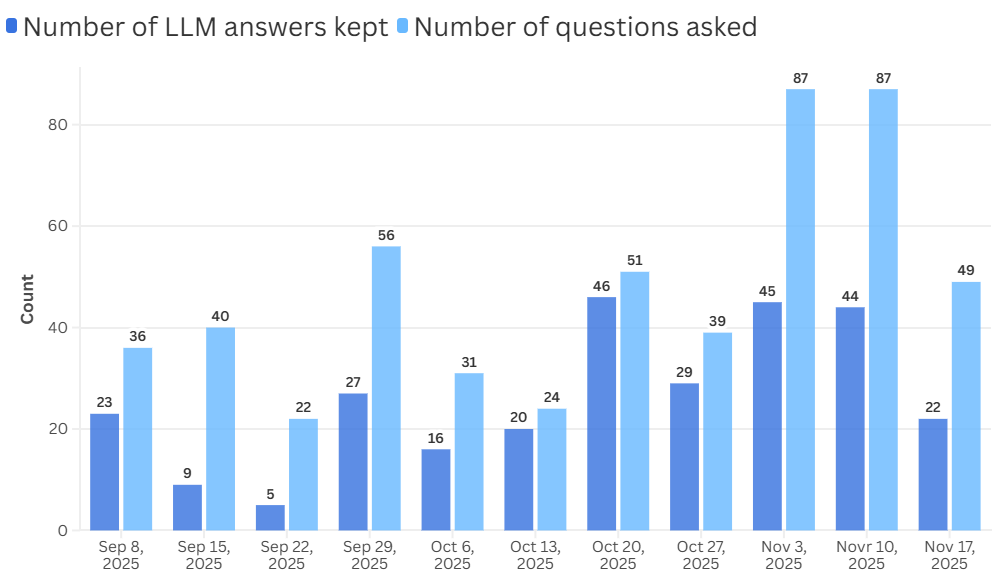}
  \caption{Weekly usage statistics of asked and kept questions}
  \Description{}
  \label{fig:combined_proportion}
\end{figure*}

\subsection{Model prompts}
The prompt used for inference is provided in Appendix \ref{app:prompts}.

\subsection{Model costs}\label{app:costs}
We have conducted the cost analysis of our pipeline relative to leading closed-source models. Under identical conditions, our system is approximately 56× more cost-effective than GPT-5 or Gemini 2.5 Pro, and 3.5× more cost-effective than DeepSeek. These estimates are based on compute prices from our university’s GPU cluster, where an A100 80GB GPU-hour is priced at \$0.37. Detailed cost estimates are presented in the Table~\ref{tab:model_prices}, costs are considered as of November 2025: \href{ https://ai.google.dev/gemini-api/docs/pricing?hl=en}{Gemini 2.5 Pro}, \href{ https://platform.openai.com/docs/pricing}{OpenAI}, \href{ https://platform.openai.com/docs/pricing}{OpenAI}, \href{ https://api-docs.deepseek.com/quick_start/pricing}{DeepSeek}.

\begin{table*}[t]
\centering
    \caption{Comparison of model pricing per 1M tokens.}
    \begin{tabular}{lcc}
    \toprule
    \textbf{Model} & \textbf{Price / 1M Input Tokens (\$)} & \textbf{Price / 1M Output Tokens (\$)} \\
    \midrule
    Our pipeline      & 0.050 & 0.15 \\
    GPT-5             & 1.25  & 10   \\
    GPT-5 mini        & 0.25  & 2    \\
    GPT-5 nano        & 0.05  & 0.40 \\
    DeepSeek          & 0.28  & 0.42 \\
    Gemini 2.5 Pro    & 1.25  & 10   \\
    \bottomrule
\end{tabular}
\label{tab:model_prices}
\end{table*}

\subsection{Generative AI use}
We used Grammarly~\citep{grammarly} to check grammar and punctuation, and GPT-4o to polish the writing, refine grammar, and improve word choice for clearer expression of ideas.






\section{Human Annotation}\label{app:Human_ann}

\subsection{Consent Form}
\begin{figure*}[t]
\centering
\begin{minipage}{0.48\textwidth}
  \centering
  \includegraphics[width=\linewidth]{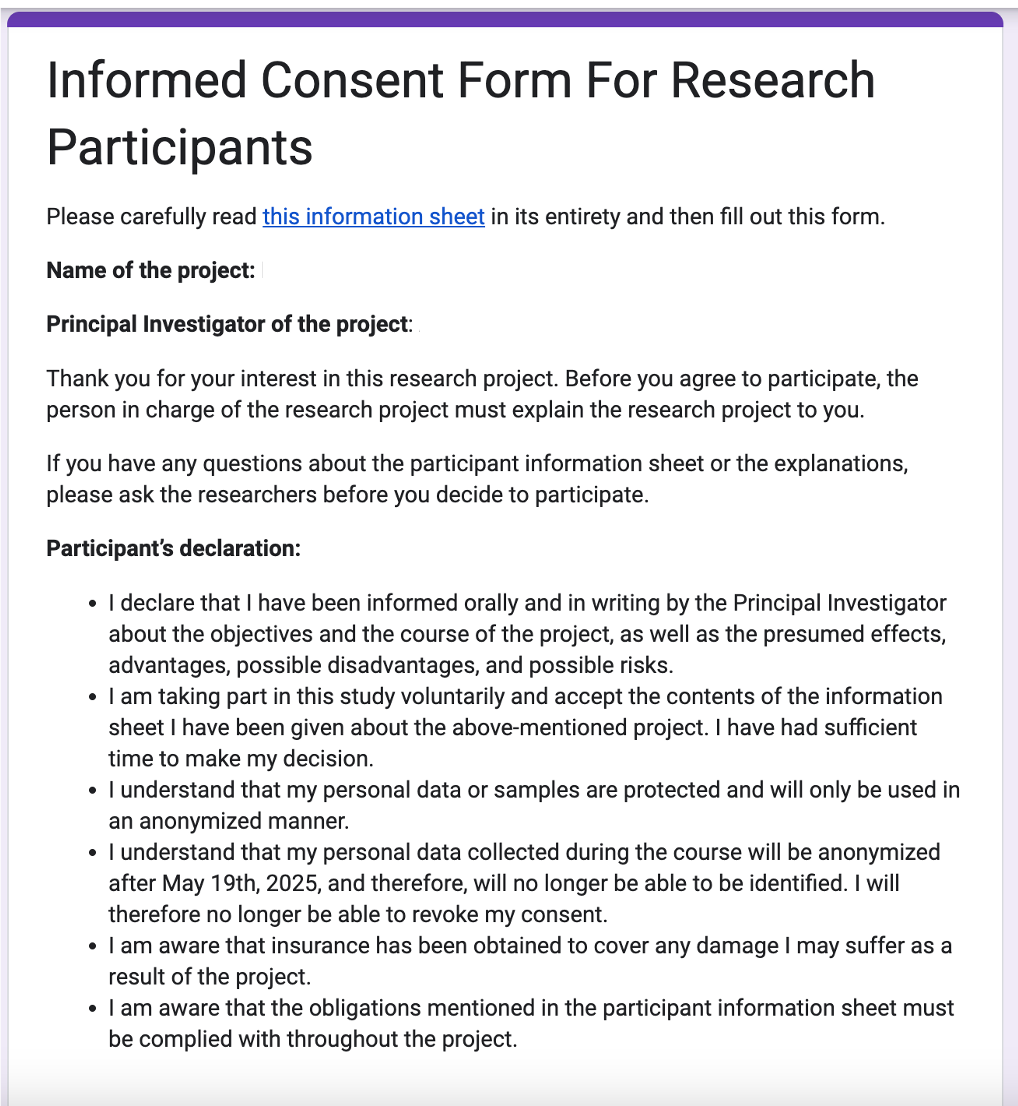}
  \caption{Consent form. Part 1.}
  \label{app:consent_1}
\end{minipage}
\hfill
\begin{minipage}{0.48\textwidth}
  \centering
  \includegraphics[width=\linewidth]{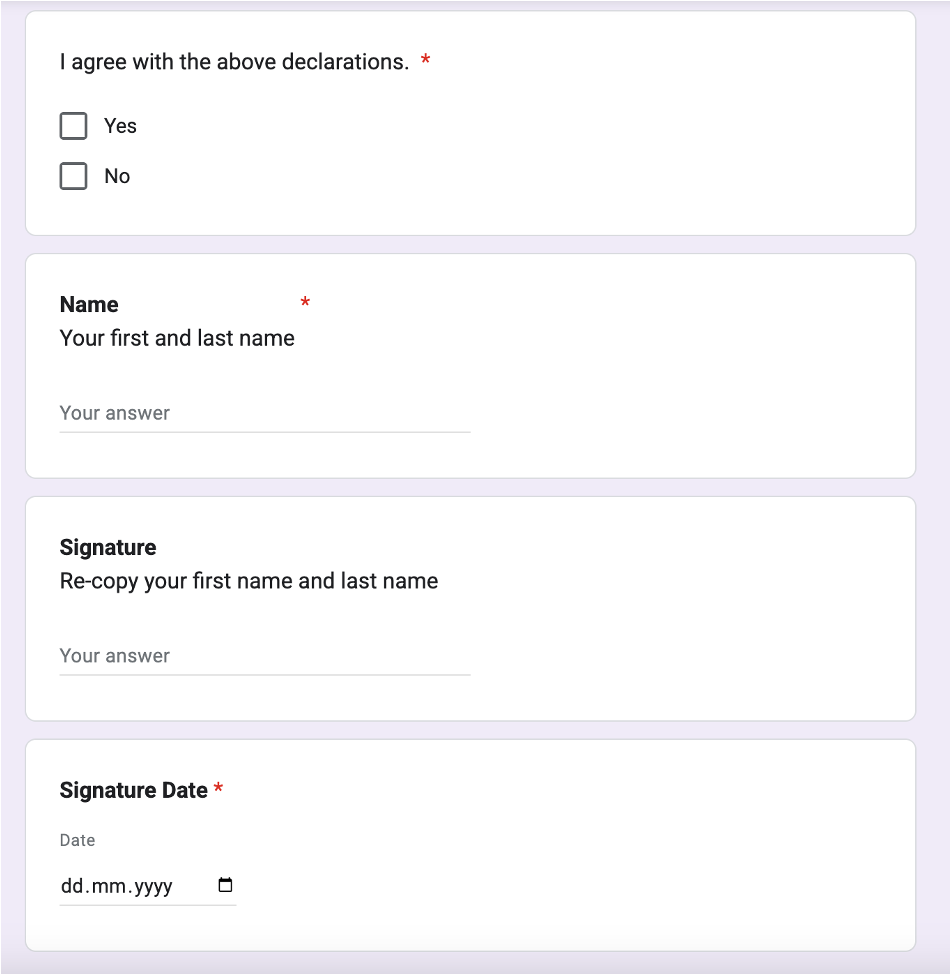}
  \caption{Consent form. Part 2.}
  \label{app:consent_2}
\end{minipage}
\end{figure*}

The consent form is provided in Figures \ref{app:consent_1} and \ref{app:consent_2}. Every participant got a copy of the consent form by email.


\subsection{Survey Layout}
\begin{figure*}[t]
  \centering
  \begin{minipage}{0.48\textwidth}
    \centering
    \includegraphics[width=\linewidth]{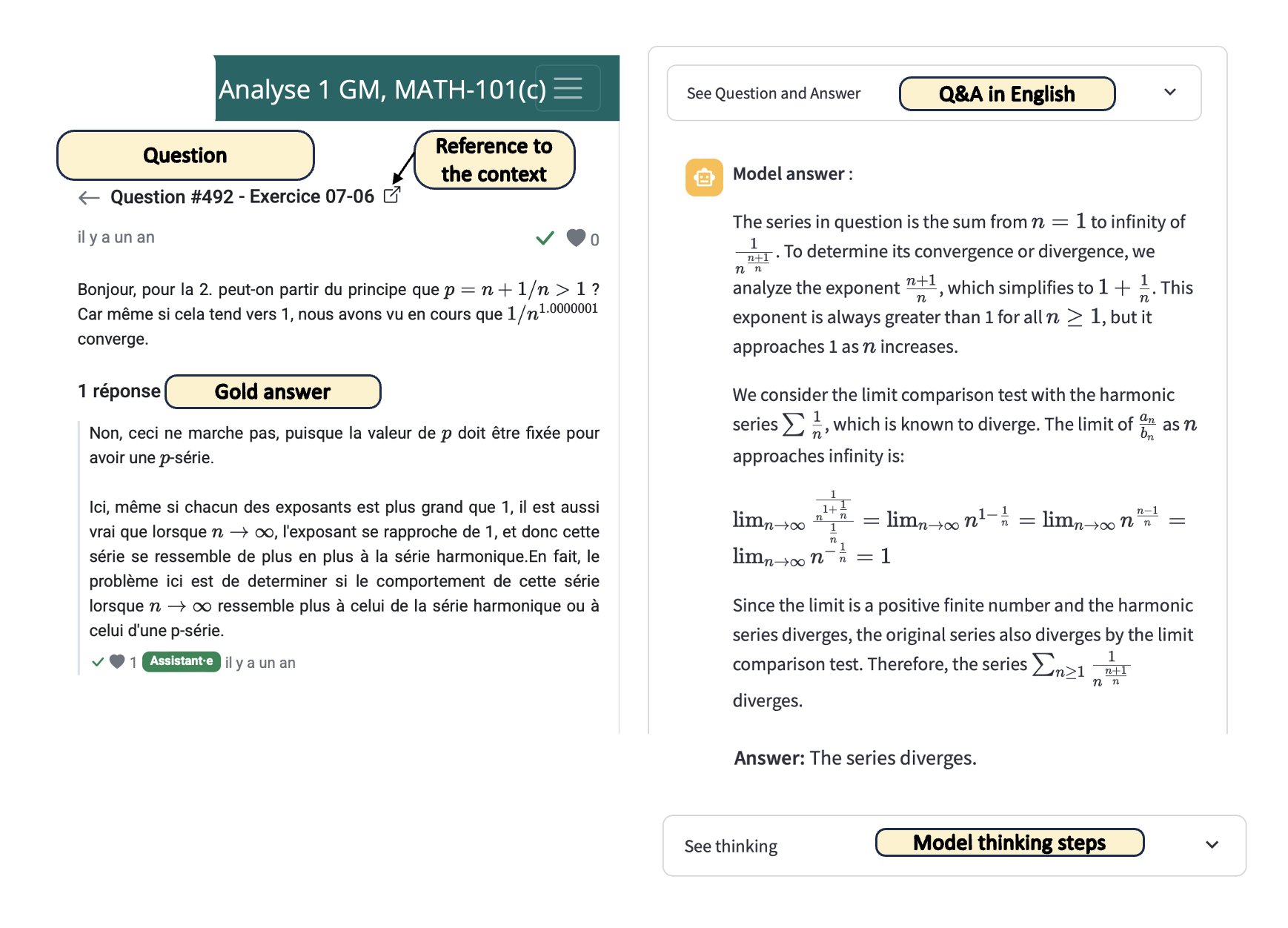}
    \caption{Survey with comments. Part 1.}
    \Description{}
    \label{app:survey_1}
  \end{minipage}
  \hfill
  \begin{minipage}{0.48\textwidth}
    \centering
    \includegraphics[width=\linewidth]{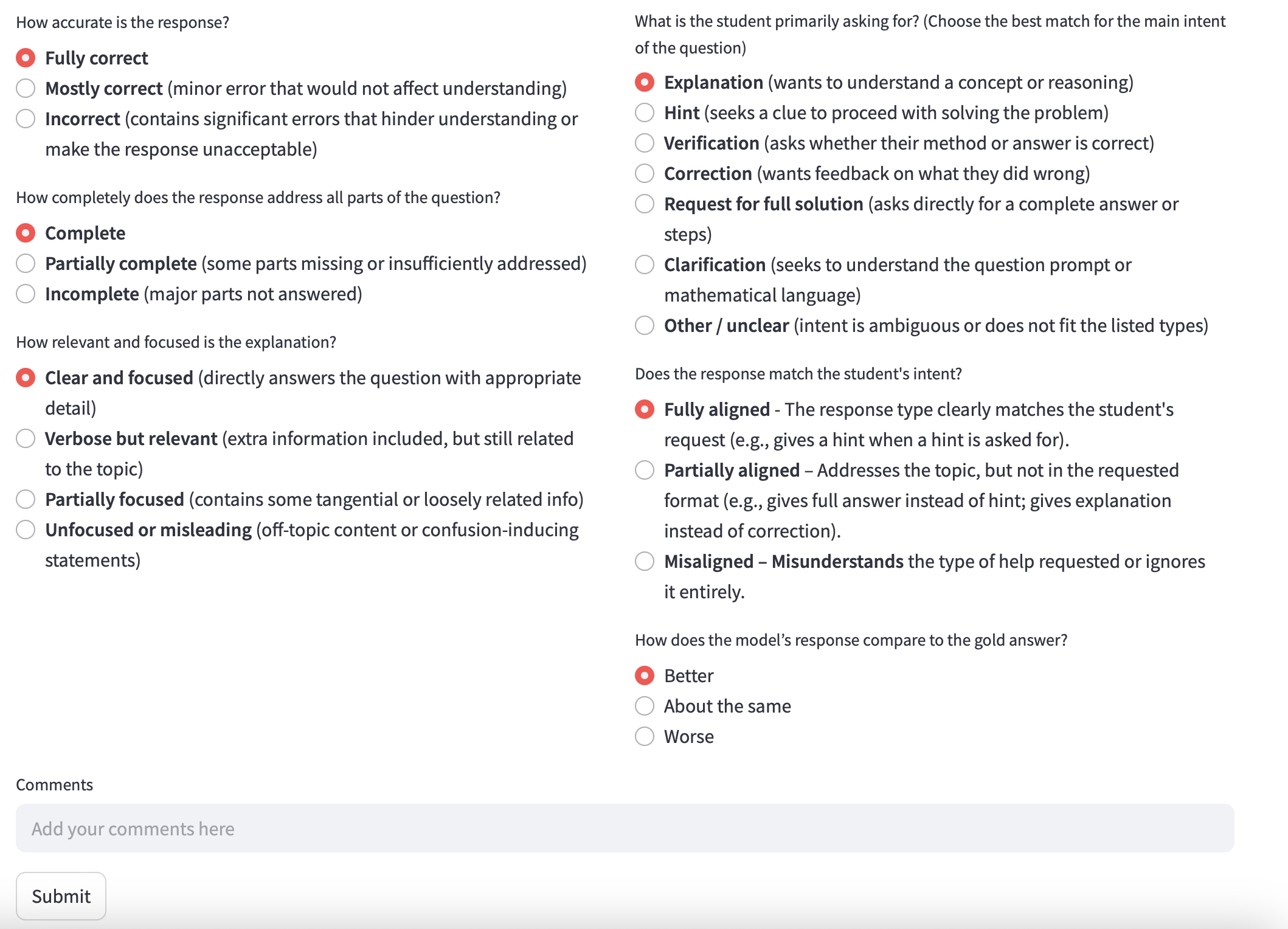}
    \caption{Survey - question part. Part 2.}
    \Description{}
    \label{app:survey_2}
  \end{minipage}
\end{figure*}

The survey layout as seen by the annotators is presented in Figures \ref{app:survey_1} and \ref{app:survey_2}.

\subsection{Student Survey}\label{app:student_survey}
\begin{figure*}[t]
\centering
\begin{minipage}{0.48\textwidth}
  \centering
  \includegraphics[width=\linewidth]{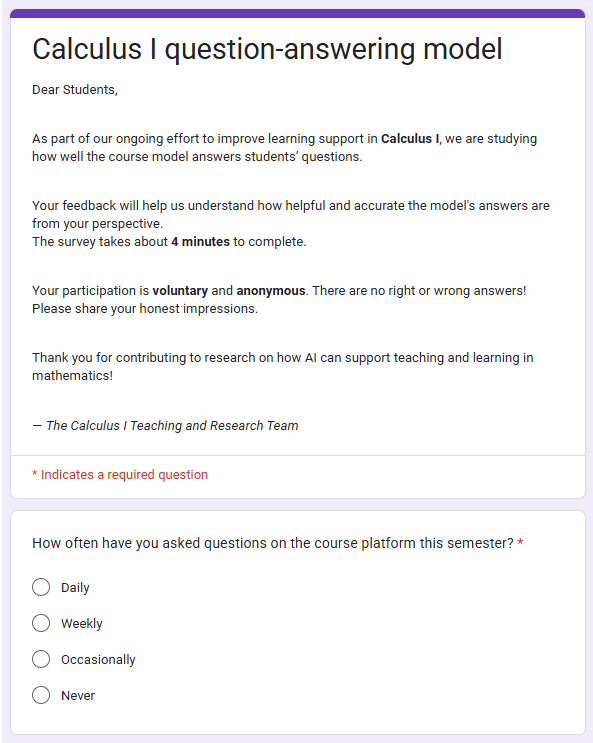}
  \caption{Student survey. Part 1.}
  \label{app:student_survey_1}
\end{minipage}
\hfill
\begin{minipage}{0.48\textwidth}
  \centering
  \includegraphics[width=\linewidth]{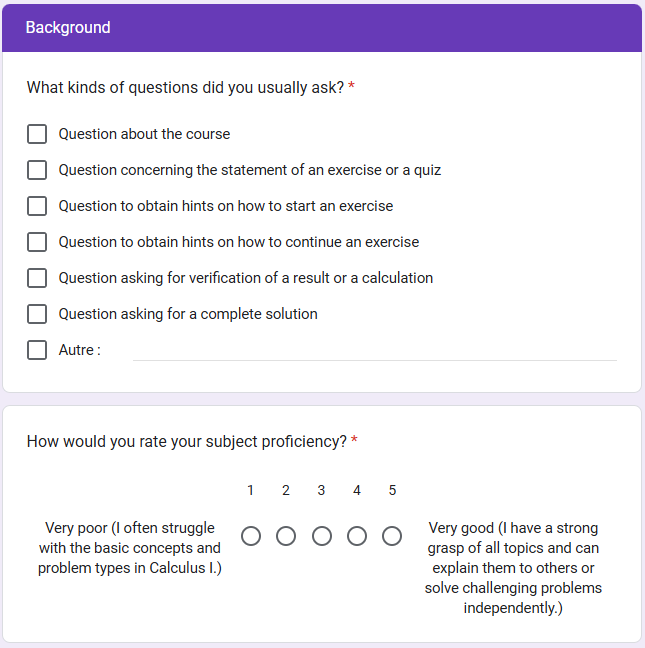}
  \caption{Student survey. Part 2.}
  \label{app:student_survey_2}
\end{minipage}
\end{figure*}

\begin{figure*}[t]
\centering
\begin{minipage}{0.48\textwidth}
  \centering
  \includegraphics[width=\linewidth]{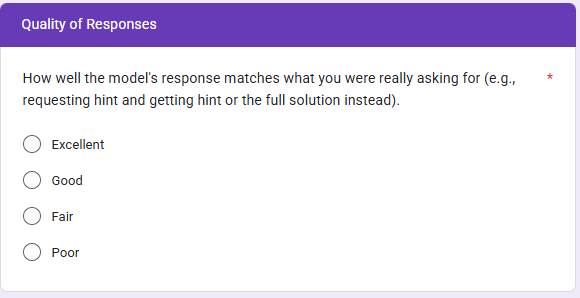}
  \caption{Student survey. Part 3.}
  \label{app:student_survey_3}
\end{minipage}
\hfill
\begin{minipage}{0.48\textwidth}
  \centering
  \includegraphics[width=\linewidth]{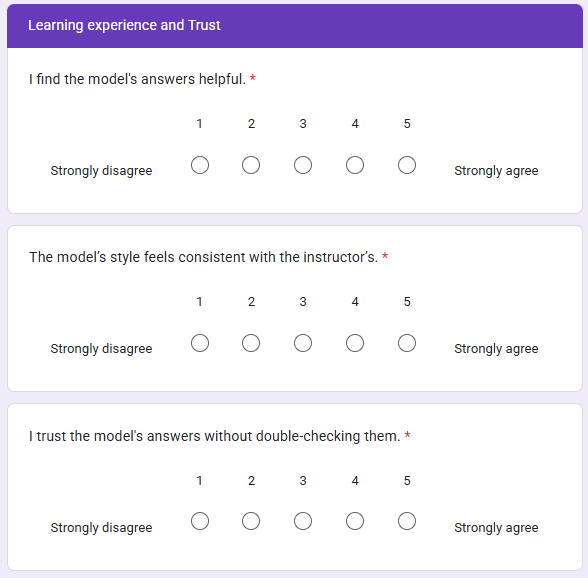}
  \caption{Student survey. Part 4.}
  \label{app:student_survey_4}
\end{minipage}
\end{figure*}

\begin{figure*}[t]
\centering
  \includegraphics[width=0.5\textwidth]{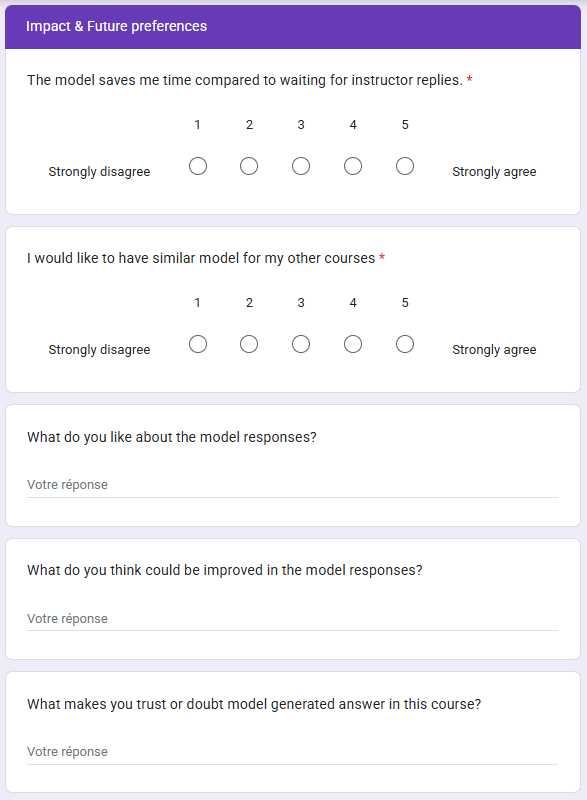}
  \caption{Student survey. Part 5.}
  \label{app:student_survey_5}
\end{figure*}

The student answer form is provided in figures \ref{app:student_survey_1}, \ref{app:student_survey_2}, \ref{app:student_survey_3}, \ref{app:student_survey_4} and \ref{app:student_survey_5}. All data collected is voluntary and anonymous, and approved by the institution's IRB.

\section{Prompts}\label{app:prompts}
\clearpage
We present the prompt template used for the answer augmentation.\\

\definecolor{promptyellow}{RGB}{255, 255, 204}
\definecolor{frameyellow}{RGB}{204, 204, 0}
\definecolor{promptblue}{RGB}{225, 240, 255}
\definecolor{frameblue}{RGB}{0, 102, 204}
\definecolor{promptpurple}{RGB}{240, 240, 255}
\definecolor{framepurple}{RGB}{75, 0, 130}
\definecolor{promptcyan}{RGB}{225, 255, 255}
\definecolor{framecyan}{RGB}{0, 153, 153}

\newsavebox{\boxcontent}

\newcommand{\CustomBox}[4]{%
    \par\noindent
    \begin{minipage}{\linewidth} %
        \colorbox{#2}{%
            \makebox[\dimexpr\linewidth-2\fboxsep][l]{%
                \color{white}\bfseries\small\hspace{1em}#1%
            }%
        }%
        \par\nointerlineskip
        \colorbox{#3}{%
            \makebox[\dimexpr\linewidth-2\fboxsep][l]{%
                \hspace{1em}%
                \begin{minipage}{\dimexpr\linewidth-2\fboxsep-2em}
                    \vspace{0.5em}
                    \usebox{#4}
                    \vspace{0.5em}
                \end{minipage}%
            }%
        }%
    \end{minipage}
    \par\vspace{1em}%
}

\begin{lrbox}{\boxcontent}
    \begin{minipage}{\dimexpr\linewidth-2\fboxsep-2em}
        \small
        \texttt{System Prompt:} Answer Calculus-related questions with a formal yet friendly tone, maintaining politeness markers to display friendliness.\\

        You are a teacher specialized in Calculus. You are provided with user questions, their related context, and a gold answer. Use your knowledge and the given context to provide the best possible answers. Be formal yet friendly and supportive to the students you wish to help.

        \begin{itemize}
            \item When the context involves an exercise with a solution, do not reveal the entire solution. Specifically, do not repeat the content within \texttt{<div id="solution" class="tab-div d-none">...</div>}. Instead, provide an answer with hints to guide the user toward finding the solution independently.
            \item Disregard the context if it is not relevant to the question.
            \item If uncertain about an answer, acknowledge your limitation politely and offer guidance, such as helping to clarify the question or suggesting relevant resources.
            \item For questions requiring heavy computations, feel free to include additional computational steps where necessary to reach a clear and accurate answer.
        \end{itemize}

        \noindent\textbf{Output Format:} \\
        Provide responses in a formal yet friendly paragraph, including politeness markers where appropriate. Hint the student to find the solution independently when applicable.

        \noindent\textbf{Notes:}\begin{itemize}
            \item Maintain a balance between formality and friendliness in your tone.
            \item Offer encouragement and courteous gestures in your responses.
            \item It is important to try the best you can in hinting the student to find the solution instead of spoiling the entire solution.
        \end{itemize}

        \vspace{0.5em}

        \noindent\textbf{Prompt Template:}
\begin{verbatim}
You are provided with a question, its context,
and the endorsed answer. Provide a new answer
that is in agreement with the gold answer, but 
feel free to give more details to 
provide a medium-sized explanation. Don't go 
too deep into the details.

CONTEXT:
{context}

QUESTION:
{question}

ANSWER:
{gold_answer}
\end{verbatim}
    \end{minipage}
\end{lrbox}

\CustomBox{Answer Augmentation Prompt}{frameyellow}{promptyellow}{\boxcontent}

\newpage
Prompt template for the translation from French to English.
\vspace{0.5em}

\begin{lrbox}{\boxcontent}
    \begin{minipage}{\dimexpr\linewidth-2\fboxsep-2em}
        \small
        Translate the provided French question and answer text into English, ensuring accuracy.

        - Retain the exact meaning of any included LaTeX, particularly with mathematical formulas. Translate only where necessary, keeping the mathematical meaning unchanged.

        - Provide only the translated text without any additional information or commentary.

        Steps:

        1. Read the French question or answer text.\\
        2. Identify and handle any LaTeX content, preserving its mathematical accuracy.\\
        3. Translate the rest of the text into clear and accurate English.\\
        4. Review the translation for accuracy and completeness.\\

        Output Format:

        Provide a clean English translation of the text, preserving the meaning and structure of any Latex elements. Do not include additional comments.

        You will be provided with .tex file that contains LaTeX commands and non-English text. Your task is to translate the non-English text into English while strictly preserving all LaTeX formatting. Follow these rules carefully:

        - Absolutely retain all LaTeX commands such as \texttt{\textbackslash begin\{defin\}}
        \texttt{\textbackslash intervalstart\{\}}
        \texttt{\textbackslash inputquiz\{\}}, and other LaTeX syntax.

        - Do not modify, remove, or reformat any LaTeX commands—the code should remain compilable.\\
        - Only translate the text content, but ensure you do not change any of the LaTeX structure or syntax.\\
        - Do not add anything extra to the translation. Focus only on converting the non-English text into English.\\
        - Ensure the entire content is translated accurately. If the input is long, translate the whole text in parts to avoid cutting off content prematurely.\\

        Translate only the text while leaving the LaTeX commands and formatting untouched. Your response should consist of the translated text with intact LaTeX formatting.
    \end{minipage}
\end{lrbox}

\CustomBox{Translation Prompts}{frameblue}{promptblue}{\boxcontent}

\clearpage
Prompt template used for inference.\\

\begin{lrbox}{\boxcontent}
    \begin{minipage}{\dimexpr\linewidth-2\fboxsep-2em}
        \small
        \texttt{System Prompt:} You are a teacher specialized in Calculus.  You are provided with user questions and their related context.  Use your knowledge and the given context to provide the best possible answers.\\

        When the context is an exercise with its solution, it is VERY IMPORTANT  that you don't reveal the whole solution. More precisely, you should not repeat everything that is contained within <div id="solution" class="tab-div d-none">...</div>, but just provide an answer with some hints that can help the user find the solution himself.\\

        If the context is not relevant to the question, disregard it. \\

        If you are unsure of the answer, politely acknowledge your limitation and offer guidance, such as clarifying the question or suggesting a related resource.\\

        Try to mainly give hints to the student and talk to him normally as a person. The hints you provide should be clear and concise, and based on how you would solve the problem yourself correctly.\\

        \texttt{Prompt Template:}\\
        CONTEXT: {context}\\

        Try to mainly give hints to the student and talk to him normally as a person. The hints you provide should be clear and concise, and based on how you would solve the problem yourself correctly.\\
        QUESTION: {question}\\
    \end{minipage}
\end{lrbox}

\CustomBox{Inference Prompt}{framepurple}{promptpurple}{\boxcontent}

Prompt template for the automated evaluation, example of correctness.
\vspace{0.5em}

\noindent

\begin{lrbox}{\boxcontent}
    \begin{minipage}{\dimexpr\linewidth-2\fboxsep-2em}
        \small
        \texttt{System Prompt:} Evaluate and compare responses to Calculus 1 forum questions with a focus on accuracy, clarity, and thoroughness.\\

        Consider each student's understanding of the calculus concept being discussed.\\

        Identify any errors or misconceptions and evaluate how well the explanations clarify these concepts.\\

        Provide constructive feedback on how the responses can be improved.\\

        \texttt{Prompt Template:}\\
        You are a Calculus 1 instructor responsible for evaluating model answers on the provided questions. Your evaluation must be rigorous, precise, and based on clear reasoning. Focus solely on the mathematical accuracy of the answer.
        Your task is to determine how correct a response is compared to the instructor's reference answer (golden answer), using the strict criteria below.\\
        ---\\
            \end{minipage}
\end{lrbox}

\CustomBox{Automated Evaluation. Part 1.}{framecyan}{promptcyan}{\boxcontent}

\newpage
\begin{lrbox}{\boxcontent}
    \begin{minipage}{\dimexpr\linewidth-2\fboxsep-2em}
        \small
        CONTEXT:
        {context}\\
        STUDENT QUESTION:
        {problem\_statement}\\
        INSTRUCTOR ANSWER (GOLDEN ANSWER):
        {golden\_answer}\\
        ANSWER TO EVALUATE:
        {generated\_answer}

        ---\\
        \# Evaluation Process\\
        Follow these steps carefully:\\
        \\
        1. **Understand the Question and Answers**\\
           - Read the student question and comprehend what a good answer should involve.\\
           - Review the instructor's (golden) answer as a reference.\\
           - Read the model answer to be evaluated.\\
        \\
        2. **Compare Answers**\\
           - Compare the model answer against the golden answer.\\
           - Focus on mathematical accuracy and conceptual correctness.\\
           - Be meticulous and critical — any error must be justified.\\
        \\
        3. **Assess Clarity and Relevance (but not verbosity)**\\
           - Do not penalize for verbosity unless it introduces confusion or includes unrelated content.\\
        \\

        4. **Score Based on the Following Criteria**\\
        \\
        **CORRECTNESS**\\
        - **2 (Fully correct)**: The answer is mathematically accurate with no conceptual or computational errors.\\
        - **1 (Mostly correct)**: The answer has minor mistakes (e.g., small arithmetic errors or irrelevant phrasing), but they do not impair understanding.\\
        - **0 (Incorrect)**: The answer has major errors, misconceptions, or omissions that affect correctness or clarity.\\
        \\
        Be objective. A low score is acceptable when justified. Grading must reflect mathematical rigor.\\
        ---\\
        Return only a JSON object with the following structure:\\
\begin{verbatim}
{
"grading_reasoning": 
"<Your reasoning here>",
"correctness": <score 0, 1 or 2>
}
\end{verbatim}
    \end{minipage}
\end{lrbox}

\CustomBox{Automated Evaluation. Part 2.}{framecyan}{promptcyan}{\boxcontent}

\end{document}